\documentclass[]{aastex631}

\usepackage{comment}
\usepackage[utf8]{inputenc}
\usepackage{amssymb}
\usepackage{amsmath}
\usepackage{hyperref}
\usepackage{tablefootnote}
\usepackage{threeparttable}

\usepackage{bibunits}
\usepackage{soul}
\begin{document}

\title{X-ray Polarization of the Black Hole X-ray Binary 4U 1630--47 Challenges Standard Thin Accretion Disk Scenario}

\author[0000-0003-0411-4243]{Ajay Ratheesh}
\affiliation{INAF Istituto di Astrofisica e Planetologia Spaziali, Via del Fosso del Cavaliere 100, 00133 Roma, Italy}
\author[0000-0003-0079-1239]{Michal Dovčiak}
\affiliation{Astronomical Institute of the Czech Academy of Sciences, Boční II 1401/1, 14100 Praha 4, Czech Republic}
\author[0000-0002-1084-6507]{Henric Krawczynski}
\affiliation{Physics Department and McDonnell Center for the Space Sciences, Washington University in St. Louis, St. Louis, MO 63130, USA}
\author[0000-0001-5418-291X]{Jakub Podgorný}
\affiliation{Université de Strasbourg, CNRS, Observatoire Astronomique de Strasbourg, UMR 7550, 67000 Strasbourg, France}
\affiliation{Astronomical Institute of the Czech Academy of Sciences, Boční II 1401/1, 14100 Praha 4, Czech Republic}
\affiliation{Astronomical Institute, Charles University, V Holešovičkách 2, CZ-18000, Prague, Czech Republic}
\author[0009-0001-4644-194X]{Lorenzo Marra}
\affiliation{Dipartimento di Matematica e Fisica, Universit\`a degli Studi Roma Tre, Via della Vasca Navale 84, 00146 Roma, Italy}
\author[0000-0002-5767-7253]{Alexandra Veledina}
\affiliation{Department of Physics and Astronomy, 20014 University of Turku, Finland}
\affiliation{Nordita, KTH Royal Institute of Technology and Stockholm University, Hannes Alfvéns väg 12, SE-10691 Stockholm, Sweden}
\author[0000-0003-3733-7267]{Valery F. Suleimanov}
\affiliation{Institut f\"ur Astronomie und Astrophysik, Universität Tübingen, Sand 1, 72076 T\"ubingen, Germany}
\author[0000-0001-5256-0278]{Nicole Rodriguez Cavero}
\affiliation{Physics Department and McDonnell Center for the Space Sciences, Washington University in St. Louis, St. Louis, MO 63130, USA}
\author[0000-0002-5872-6061]{James F. Steiner}
\affiliation{Center for Astrophysics | Harvard \& Smithsonian, 60 Garden St, Cambridge, MA 02138, USA}
\author[0000-0003-2931-0742]{Ji\v{r}\'{i} Svoboda}
\affiliation{Astronomical Institute of the Czech Academy of Sciences, Boční II 1401/1, 14100 Praha 4, Czech Republic}
% ...



\begin{abstract}
Large energy-dependent X-ray polarization degree is detected by the Imaging X-ray Polarimetry Explorer ({IXPE}) in the high-soft emission state of the black hole X-ray binary 4U 1630--47. The highly significant detection (at $\approx50\sigma$ confidence level) of an unexpectedly high polarization, rising from $\sim6\%$ at $2$ keV to $\sim10\%$ at $8$ keV, cannot be easily reconciled with standard models of thin accretion discs. In this work we compare the predictions of different theoretical models with the {IXPE} data and conclude that the observed polarization properties are compatible with a scenario in which matter accretes onto the black hole through a thin disc, covered by a partially-ionized atmosphere flowing away at mildly relativistic velocities.
\end{abstract}

\keywords{Polarimetry (1278) --- X-ray astronomy (1810) --- Stellar mass black holes (1611) -- Accretion (14)}

\section{Introduction} \label{sec:intro}

Black hole X-ray binaries (BHXRBs) 
consist
of a black hole (BH) accreting matter from a companion star. 
These systems provide opportunities to investigate the inner workings of accretion flows, including their thermal stability and the mechanisms of angular momentum transport, as well as the formation of relativistic outflows, winds and jets \citep{Ponti_2012,Meszaros_1997, Fender_2001}.

The X-ray polarization observations (in the 2--8 keV range) conducted with the Imaging X-ray Polarimetry Explorer \cite[{IXPE},][]{Weisskopf_2022}, a NASA mission in collaboration with the Italian Space Agency (ASI), launched on 2021 December 9, are providing us with new significant insights into BHXRBs. 
In particular, the 4\% polarization observed for the archetypal BHXRB Cyg X-1 \citep{Krawczynski_2022}, parallel to the radio jet, constrained the geometry of the hot corona responsible for the power-law emission in the hard state, to be extended perpendicular to the jet direction
\cite[see][for detailed descriptions of the different X-ray states]{Remillard_2006,2007A&ARv..15....1D}. IXPE monitored the polarization properties of the recently discovered black hole binary candidate, Swift J1727.8$-$1613, throughout its state transition from a hard to a soft state, suggesting that the comptonizing corona extends over the disk, similar to the case of Cyg X-1 \citep{Veledina_2023b,Ingram_2023}. In the disk-dominated spectral states of LMC X-1, LMC X-3, and 4U 1957+115, IXPE observations indicates that the polarization measurements are consistent with the standard geometrically-thin optically-thick accretion disks \citep{Podgorny_2023, Svoboda_2023, Marra_2023}. Furthermore, in case of LMC X-3 and 4U 1957+115, these observations place constraints on the spin of the black hole. On the other hand, the much higher polarization degree ($\sim20\%$) observed for the BH candidate Cyg X-3 in the hard state \citep{Veledina_2023}, with polarization direction perpendicular to that of the discrete radio blobs, can be explained if radiation emitted by a highly luminous (but not directly visible) source is reflected off the funnel walls of obscuring matter.

In this paper we present the results 
of spectro-polarimetric observations of the accreting BH 4U\,1630--47 with {IXPE}, accompanied by spectral observations with the {NICER} and {NuSTAR} missions, discussing a comparison with the predictions of different theoretical models. 
IXPE observed 4U\,1630--47 in the high-soft state, in which the emission is believed to originate from a geometrically thin, optically thick accretion disk ($h/r\ll 1$ for the scale height $h$ of the disk at radius $r$), manifesting itself as a multi-temperature black-body \cite[BB,][]{Shakura_1973,Novikov_1973,Esin_1997,MR2006}. However, also slim ($h/r\lesssim 0.4$) and thick ($0.4 < h/r \lesssim 1$) accretion discs have been provided as possible interpretations in the case of high accretion rates \citep{Abramowicz_1988}.
The {IXPE} observations of 4U~1630--47 now allow us to weigh in on this distinction thanks to the additional information encoded in the energy resolved polarization degree and polarization direction.

4U 1630--47 is a transient low-mass X-ray binary (LMXB) system, initially discovered by the Uhuru satellite in 1969 \citep{Giacconi_1972, Priedhorsky_1986}, that subsequently exhibited recurrent outbursts with a
spacing
of approximately 2--3 years \citep{Kuulkers_1998, Capitanio_2015}.
An accurate determination of the properties of the binary system has not been possible as yet due to the high line-of-sight (LOS) extinction \citep{Reid_1980, Parmar_1986}. 
The BH mass, distance to the binary system and inclination (angle between the binary axis and the LOS) are thus poorly constrained. 
Based on the dust scattering halo around the source, the distance is estimated to be between $4.7$ to $11.5$ kpc  \citep{Kalemci_2018}, while 
the inclination is believed to be $\sim65\degr$, explaining the observations of X-ray dips but the absence of eclipses 
\citep{Tomsick_1998,Kuulkers_1998} and the
detection of Doppler shifted lines emitted by a jet \citep{Trigo_2014}. 
The source shows evidence for a wind, believed to be equatorial
\citep{Trigo_2014, King_2014, Miller_2015}. 
The thermal component usually dominates the spectrum during outbursts \citep{Parmar_1986, Tomsick_2005}, making it an ideal candidate for investigating the properties of the disc.

We present the observational results in Section \ref{results}, and the theoretical modeling in Section \ref{sec:theormodels}. In Section \ref{sec:comparisonmodeldata} we compare the models to the data and conclude by discussing the implication of the results in Section \ref{discussion}.

\section{Observational Results}\label{results}
\begin{figure}[t]
    \centering
    \includegraphics[width=8cm]
    {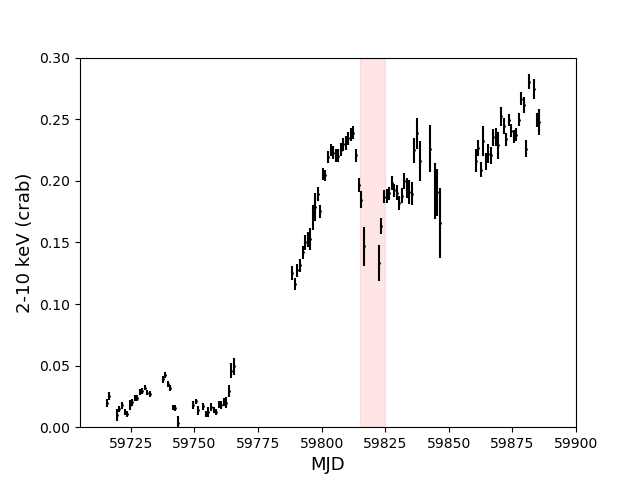}
    \includegraphics[width=8cm]{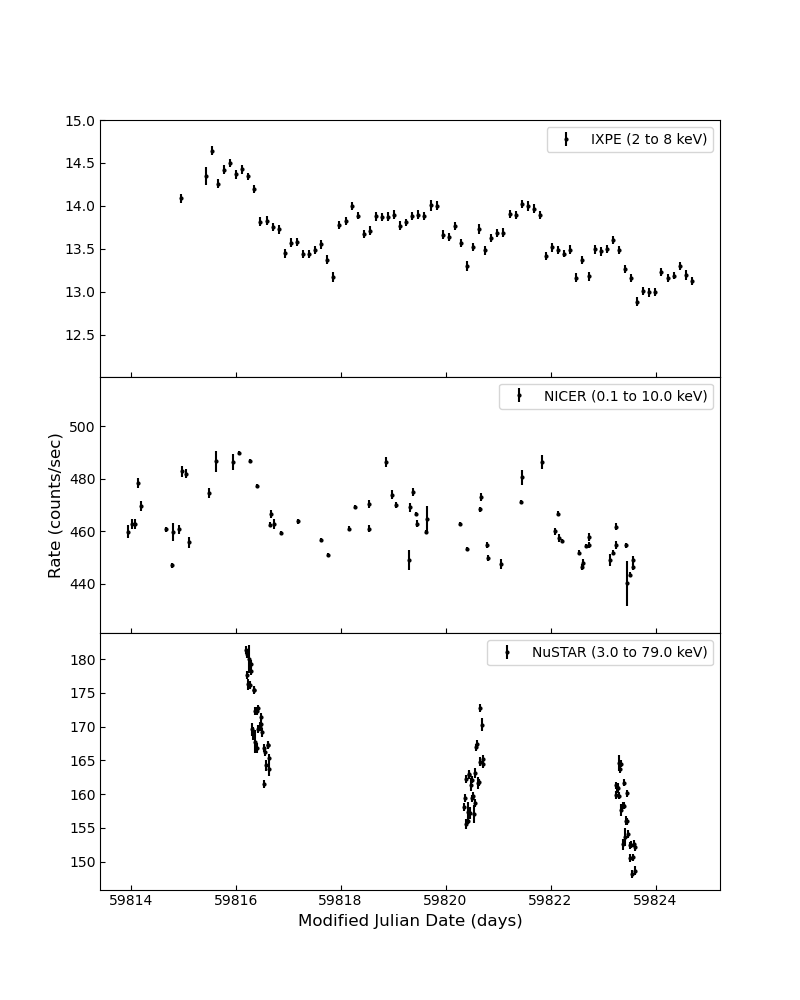}
\caption{Left: {MAXI}  2--10 keV light curves in one-day bins. The shaded red region indicates the time period of the {IXPE} observation campaign. Right: {IXPE}, {NICER} and {NuSTAR} lightcurves in the 2--8 keV, 0.1--10 keV and 3--79 keV bands, respectively, during the {IXPE} campaign. The {IXPE} light curve gives the combined count rates of all the three DUs, whereas for {NuSTAR} the light curve obtained from the Focal Plane Module A only is shown. 
}
    \label{fig:lc}
\end{figure}

Daily monitoring of 4U 1630--47 by the Gas Slit Camera (GSC) onboard Monitor of All-sky X-ray Image \cite[{MAXI},][]{Maxi_2009} showed an increase in the count rate, suggesting an outburst from the source in 2022 July \cite[][see also Figure~\ref{fig:lc}]{Jiang_2022}. During this outburst, {IXPE} performed a target of opportunity (ToO) observation starting on 2022 August 23 and ending on 2022 September 2, for a total exposure of approximately 460 ks, along with continuous spectral monitoring from the Neutron Star Interior Composition Explorer \cite[{NICER},][]{NICER_2014} and the Nuclear Spectroscopic Telescope Array \cite[{NuSTAR},][]{Nustar_2013}. In this section we show the light curves, polarimetric and spectroscopic analysis of the source.\\

{\it \bf Light curves:} The left panel of Figure~\ref{fig:lc} shows the X-ray activity of the source between MJD 59700 (2022 May 1) and MHD 59900 (2022 November 17) monitored by the {\it MAXI} mission. The 200-day interval includes the time period of the {IXPE} observation campaign. The source was detected in the high-soft state. The right panel shows the {IXPE} 2--8 keV, {NICER} 1--10 keV, and {NuSTAR} 3--79 keV fluxes. The fluxes varied by $\sim$10\% below 10 keV and by 15--20\% above 10 keV. Hence, the flux of the source was rather stable during the campaign (red shaded region).\\

\begin{figure}[h]
    \centering
\includegraphics[height=10cm]{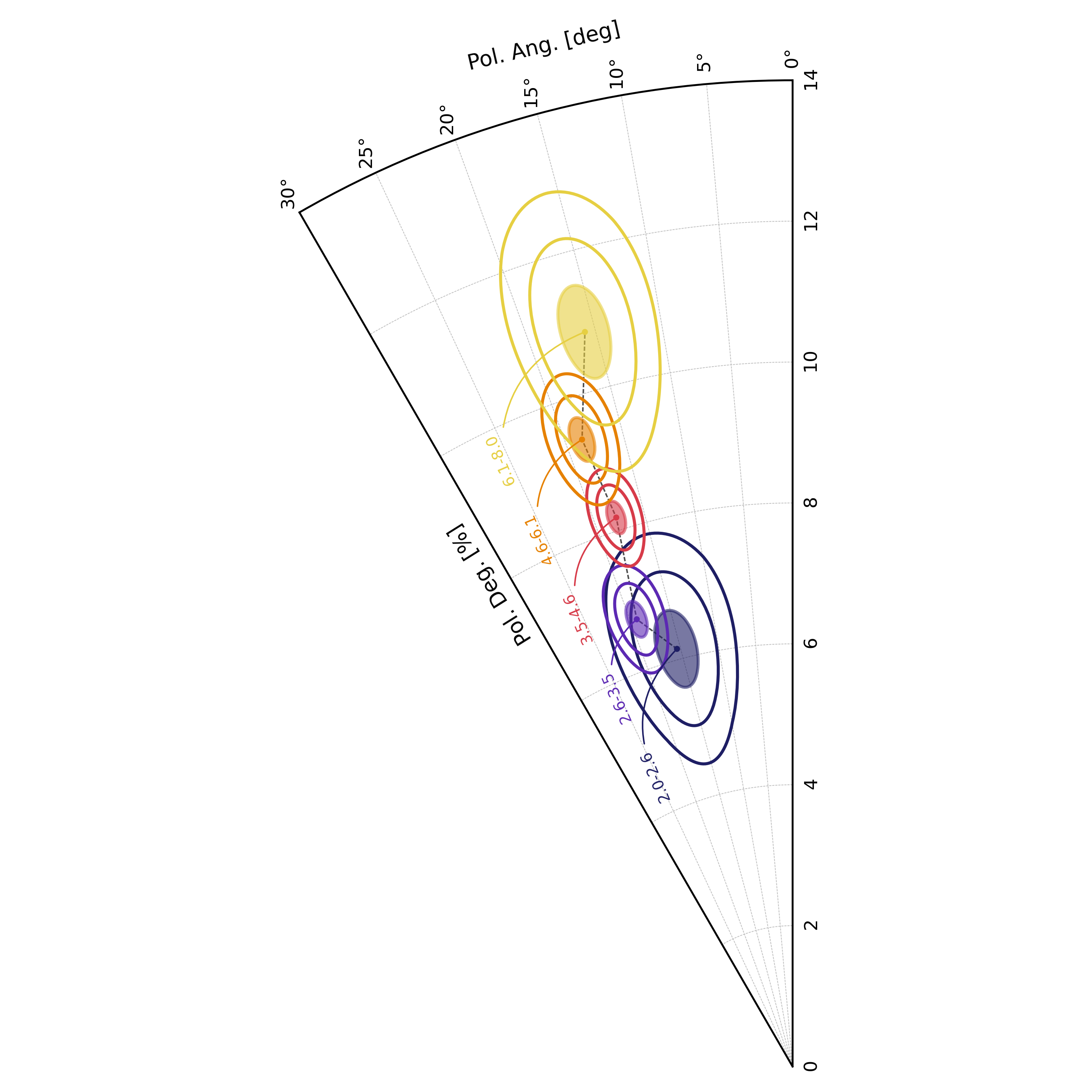}
\caption{Polarization degree and angle measured for 4U~1630--47 as functions of the energy. The analysis is carried out using the publicly available version of {\tt ixpeobssim} \citep{Baldini2022}. 
    The shaded region and the ellipses show the 68\% and the 95\% and 99.7\% confidence contours, respectively.}
    \label{fig:polarplot}
\end{figure}

{\bf Polarization}: Linear polarization was detected with a statistical confidence of around 50$\sigma$ (Figure~\ref{fig:polarplot}). The 2--8\,keV polarization degree (PD) and polarization angle (PA), measured east of North, are $8.32\pm0.17\%$ and $17.8\degr\pm 0.6\degr$, respectively (uncertainties at 68\% confidence level). Whereas PD increases from approximately 6\% at 2 keV to 10\% at 8\,keV, PA stays constant with energy within statistical accuracy (see the Table~\ref{table:pol_vs_ene}).
The radio jet from this source has never been resolved, leaving us without a jet direction to compare with the X-ray PA.  The data reduction and analysis techniques are outlined in Appendix \ref{section:datareduction}.\\

\begin{table}[b]
\centering
\caption{Polarization degree and angle in different energy bins across the {IXPE} $2$--$8$ keV band.}
\begin{tabular}{c c c} 
\hline
\hline
Energy range (keV) & PD (\%) & PA ($\degr$) \\
\hline 
2.0--2.5 & $6.1\pm0.7$ & $14.6\pm3.3$ \\
2.5--3.0 & $6.1\pm0.4$ & $17.8\pm1.9$ \\
3.0--3.5 & $7.1\pm0.3$ & $19.8\pm1.3$ \\
3.5--4.0 & $7.9\pm0.3$ & $17.8\pm1.2$ \\
4.0--4.5 & $8.7\pm0.4$ & $18.3\pm1.2$ \\
4.5--5.0 & $8.0\pm0.5$ & $16.3\pm1.7$ \\
5.0--5.5 & $10.0\pm0.5$ & $18.8\pm1.5$ \\
5.5--6.0 & $10.4\pm0.7$ & $18.4\pm1.9$ \\
6.0--6.5 & $10.1\pm0.8$ & $19.3\pm2.3$ \\
6.5--7.0 & $12.1\pm1.2$ & $12.8\pm2.7$ \\
7.0--7.5 & $9.3\pm1.6$ & $14.3\pm4.9$ \\
7.5--8.0 & $11.8\pm2.5$ & $18.9\pm6.0$ \\
\hline 
\end{tabular}
\label{table:pol_vs_ene}
\end{table}

{\bf Spectral fit with {NICER} and {NuSTAR}:} \label{section:specdataanalysis}
The {NICER} data reveal a featureless power density spectrum with fractional variability of less than $5\%$ Root Mean Square (RMS) and a high disk temperature of $\sim1.4$ keV, showing that the source was in the high-soft state. The {NICER} energy spectra also exhibit blue-shifted absorption lines, suggesting the existence of equatorial disk wind outflows with velocities $\approx0.003\,c$ (with $c$ the speed of light).\\


As 4U\,1630$-$47 is variable (see Figure~\ref{fig:lc}), we fitted the {NICER} and {NuSTAR} energy spectra selecting quasi-simultaneous data sets. From the set of {NICER} observations, we used data with observational IDs 5501010104, 5501010108 and 5501010111, that match the same periods covered by {NuSTAR} observations with observational IDs 80802313002, 80802313004 and 80802313006, respectively. Henceforth, we denote these 3 observations as Obs 1, Obs 2 and Obs 3, respectively. For the spectral analysis, we used only {NICER} data with low background radiation (we avoided here data during the SAA passage).\\

\begin{figure}[thb]
    \centering
    \includegraphics[width=0.49\textwidth, height=0.6\textwidth]{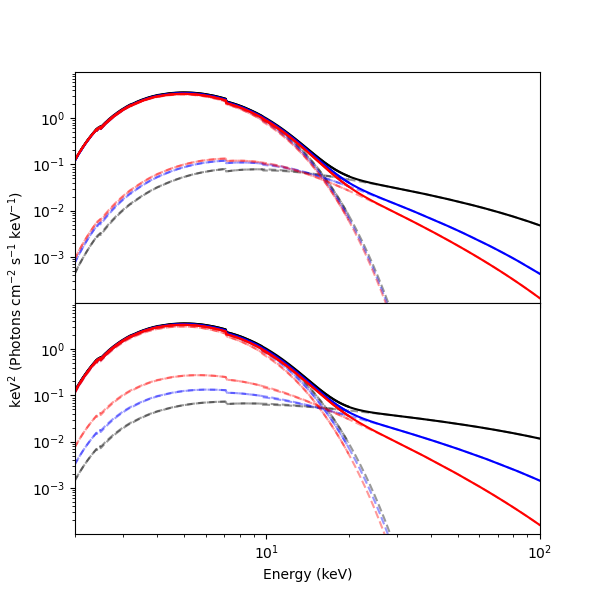}
    \includegraphics[width=0.49\textwidth, height=0.6\textwidth]{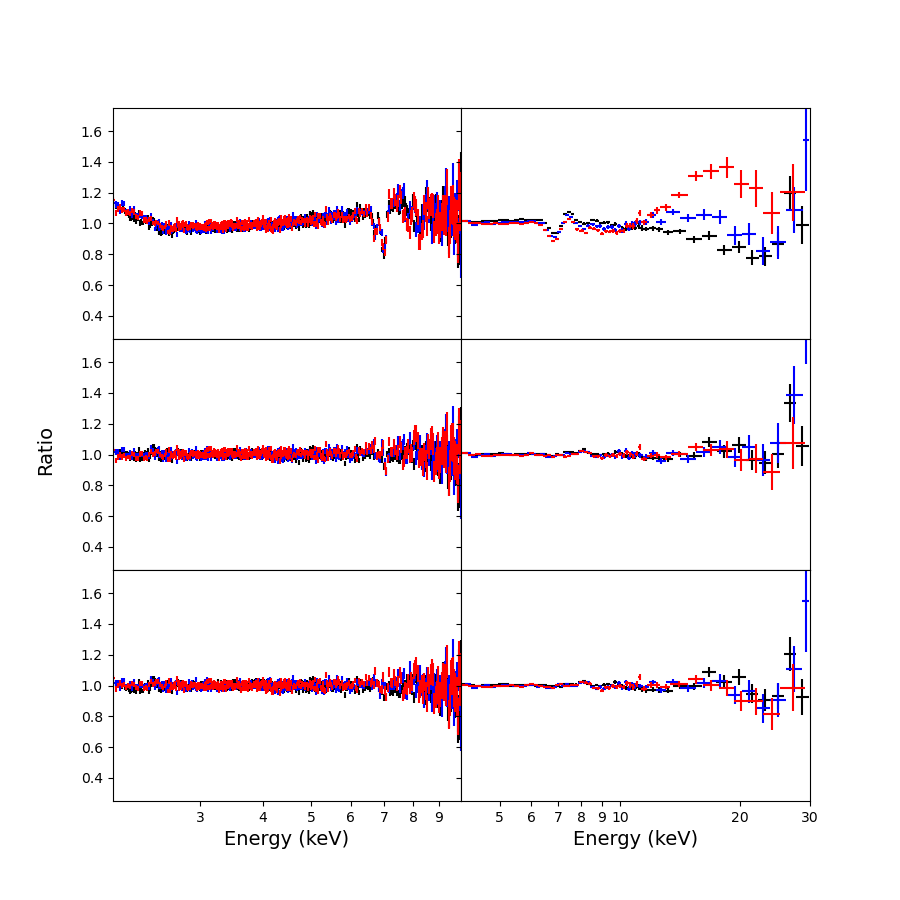}
    \caption{Left: The best fit continuum models to the {NICER} and {NuSTAR} data, using thermal accretion disk models, \texttt{kerrbb} (top) and \texttt{slimbh} (bottom), along with the comptonization component (\texttt{nthcomp}). The solid lines indicate the total model, while the dashed lines show the disk emission and comptonization components. 
    Right: The residuals (data/model) of the spectral fit NICER (left) and NuSTAR (right) data. The top, middle and bottom panels represents the residuals from SM1, SM2, and SM3, respectively. Colors black, blue, and red correspond to Obs 1, Obs 2, and Obs 3, respectively.}
    \label{fig:simult_nicernustar}
\end{figure}

Spectral Model 1 (SM1): Initiallially we fit the {NICER} and {NuSTAR} spectra using a multi-temperature black-body ({\tt diskbb}) and a power-law ({\tt po}), with galactic cold absorption ({\texttt{tbabs}}), we further used a constant to account for the calibration uncertainties between instruments. In XSPEC notation this becomes,
{\tt constant$\times$tbabs$\times$(diskbb+po)}. The power-law component was allowed to vary independently between observations. The fit resulted in an inner disk temperatures of 1.38 keV and photon indices of 3.9, 4.1, and 4.3 for obs 1, obs 2, and obs 3, respectively. However, the $\chi^2$/dof of the fit was 14692/2952, and hence not an acceptable fit. The fitted model and the fit residuals are shown in Figure \ref{fig:simult_nicernustar}\\

Spectral Model 2 (SM2): We then used a model consisting of thermal accretion disk emission accounting for relativistic effects \cite[{\texttt{kerrbb}},][]{Li_2005}, the Comptonized emission {\texttt{nthcomp}} \citep{Zdziarski1996, Zycki1999}, ionized absorber modeled with {\tt cloudy} \citep{Ferland2017} and cold absorption {\texttt{tbabs}} \citep{Wilms_2000}, accounting for Galactic as well as local absorption. The {\tt cloudy} absorption table reproduces the absorption lines self consistently through a slab with a constant density of $10^{12}$ cm$^{-3}$ and a turbulence velocity of 500 km s$^{-1}$, illuminated by the unabsorbed intrinsic best fit SED described below. Modeling the absorption lines requires a highly ionized outflowing plasma, i.e. with ionization parameter $\xi$ $\sim$ 10$^{5}$ and column density $N_\mathrm{Heq}\sim10^{24}\,{\rm cm}^{-2}$. The seed photons for the {\texttt{nthcomp}} model are assumed to be from the multicolor-BB disk emission (inp\_type parameter = 1) and we fixed the temperature to $kT_{\rm BB} = 1.38$ keV, as resulted from initial fit (SM1). We further added an empirical {\tt{edge}} model to account for the instrumental features at $\approx 2$--$3$\,keV in the {NICER} spectra and at $\approx 10$\,keV in the {NuSTAR} spectra. The gold M edge in the {NICER} spectra is a well known instrumental feature. We fixed the energy of the edge to the value $E = 2.4$\,keV, as reported by \cite{Wang2021} in the analysis of MAXI J1820+070 {NICER} spectra. The origin of the $\approx 10$\,keV edge in the {NuSTAR} spectra is less known, but it was observed in other sources as well \cite[e.g. the analysis of LMC X-1,][]{Podgorny_2023}. We, therefore, modeled it with the empirical {\tt{edge}} model. Finally, we accounted for absolute calibration uncertainties between {NICER} and {NuSTAR} instruments, allowing for a small discrepancy in the spectral slope and normalization between the instruments using the {\tt{mbpo}} model \cite[see][]{Krawczynski_2022}. This model can be denoted in XSPEC as {\tt mbpo$\times$edge$\times$tbabs$\times$cloudy$\times$(kerrbb+nthcomp)}. The fitted model and the fit residuals are shown in Figure \ref{fig:simult_nicernustar}.

We further added a reflection component, a second Comptonisation component, or a second ionized absorption component to the model, but none of these improved the overall fit or the residuals in this part of the {NuSTAR} spectrum.

With this model, we obtained the best fits for BH spins $a \gtrsim 0.99$, inclinations $i \approx 85\degr$, and masses $M_{\textrm{bh}} \gtrsim 50 M_\odot$ (Table \ref{table:nicernustar_spinincl}, last line), with $\chi^2$/dof = 3563/2937. 
The mass is much higher than that ($\approx 10\,M_\odot$) estimated from the previous analysis by \citet{Seifina2014}. 
\begin{table}[t]
\centering
\caption{Spectral fit results for different fixed values of BH spin and inclination with the {\tt kerrbb} model.}
\begin{tabular}{ccccc}
\hline
\hline
\rule{0cm}{0.3cm}
Inclination & Spin & Mass & Accretion rate & Fit goodness\\
\hline
$i$ (deg) & $a$ & $M_{\textrm{bh}}$ ($M_\odot$) & $M_{\textrm{dd}}$ ($10^{18}$\,g s$^{-1}$) & $\chi^2$ (2935 dof)\\
\hline
$70$ & $0.7$ & $9.98 \substack{+0.06 \\ -0.08}$ & $3.9-4.3$ & $4693$ \\
     & $0.998$ & $29.8\substack{+0.02 \\ -0.02}$ & $0.99-1.09$ & $3585$ \\
$85$ & $0.7$ & $16.1 \substack{+0.01 \\ -0.01}$ & $8.3-8.9$ & $3711$ \\
     & $0.998$ & $59.8\substack{+0.04 \\ -0.02}$ & $1.02-1.12$ & $3563$ \\
     \hline
\end{tabular}
\label{table:nicernustar_spinincl}
\end{table}

\begin{table}[t]
\centering
\footnotesize
\caption{Spectral fit parameters to simultaneous {NICER} and {NuSTAR} observations with the \texttt{slimbh} model.}
\resizebox{\textwidth}{!}{
\begin{tabular}{cccccc}
\hline
\hline
\rule{0cm}{0.3cm}
Comp. & Parameter (unit) & Description & Obs 1 & Obs 2 & Obs 3\\  
\hline
\rule{0cm}{0.3cm}
{\tt TBabs} & $N_\textrm{H}$ ($10^{22}$\,cm$^{-2}$) & H column density & $7.92\substack{+0.07 \\ -0.02}$ & $7.94\substack{+0.02 \\ -0.02}$ & $7.85\substack{+0.02 \\ -0.02}$ \\
\hline
\rule{0cm}{0.3cm}
\textsc{cloudy} & $\log \xi$ & ionization & $5.13\substack{+0.06 \\ -0.04}$ & $5.01\substack{+0.09 \\ -0.03}$ & $4.95\substack{+0.07 \\ -0.04}$ \\
& $\log N_\mathrm{Heq}$ & H column density & $24.03\substack{+0.02 \\ -0.02}$ & $24.03\substack{+0.03 \\ -0.01}$ & $24.04\substack{+0.03 \\ -0.02}$ \\
& $v$ (km\,s$^{-1}$) & outflow velocity & $<90$ & $-900\substack{+300 \\ -300}$ & $-900\substack{+300 \\ -300}$ \\
\hline
\rule{0cm}{0.3cm}
{\texttt{slimbh}} & $M_{\textrm{bh}}$ ($M_\odot$) & Black hole mass & \multicolumn{3}{c}{$18.0\substack{+0.7 \\ -1.2}$} \\ 
    & $a$ & Black hole spin & \multicolumn{3}{c}{$0.71\substack{+0.03 \\ -0.14}$} \\
    &  $L_{\rm Edd}$ & Luminosity & $0.53\substack{+0.03 \\ -0.03}$ & $0.51\substack{+0.02 \\ -0.02}$ & $0.49\substack{+0.02 \\ -0.02}$ \\
    & $i$ (deg) & Inclination & \multicolumn{3}{c}{$85\substack{\rm{f} \\ - 1.4}$} \\
    & $\alpha$ & Viscosity & \multicolumn{3}{c}{$0.1$ (frozen)} \\
    & $D_\textrm{bh}$ (kpc) & Distance & \multicolumn{3}{c}{$11.5$ (frozen)} \\
    & hd & Hardening factor & \multicolumn{3}{c}{$-1$ (frozen)} \\
    & $l_\textrm{flag}$ & Limb-darkening & \multicolumn{3}{c}{$0$ (frozen)} \\
    & $v_\textrm{flag}$ & Self-irradiation & \multicolumn{3}{c}{$0$ (frozen)} \\
    & norm & normalization & \multicolumn{3}{c}{$1$ (frozen)}  \\
\hline
\rule{0cm}{0.3cm}
{\texttt{nthcomp}} & $\Gamma$ & Photon index &
$2.6\substack{+0.2 \\ -0.2}$ & $3.6\substack{+0.2 \\ -0.2}$ & $4.5\substack{+0.2 \\ -0.2}$\\
& $kT_\textrm{e}$ (keV) & Electron temp. &
\multicolumn{3}{c}{$500$ (frozen)} \\
& $kT_\textrm{BB}$ (keV) & Seed photon temp. &
\multicolumn{3}{c}{$1.47$ (frozen)} \\
& norm (10$^{-2}$) & normalization &
$2.6\substack{+1.1 \\ -0.7}$ & $6.3\substack{+1.8 \\ -1.7}$ & $13\substack{+3 \\ -3}$\\
\hline
\rule{0cm}{0.3cm}
$\chi^2$ / dof & & & \multicolumn{3}{c}{3494/2933} \\
\hline
\end{tabular}
}
\label{table:nicernustar}
\begin{tablenotes}
\item {Note:} The final model also included cross-calibration uncertainties between {NICER}, {\it NuSTAR A} and {\em B} instruments, modeled by an {\tt{mbpo}} model with $\Delta\Gamma = -0.101 \pm 0.008$ and normalization $N_{\rm mbpo} = 1.16 \pm 0.02$ (consistent for both {\it NuSTAR A} and {\it B} instruments), and the instrumental edges at $E = 2.4$\,keV with maximum $\tau = 0.074 \pm 0.005$ for {NICER}, and $E = 9.7 \pm 0.1$\,keV with maximum $\tau = 0.056 \pm 0.006$ for {NuSTAR}. See the main text for more details.\\
\end{tablenotes}
\end{table}

As the BH mass, spin, inclination and distance in the {\texttt{kerrbb}} model are degenerate, we fixed the distance of the source to $D_{\rm bh} = 11.5$\,kpc \citep{Kalemci_2018}. 
A smaller $D_{\rm bh}$ of 4.7\,kpc is not excluded and would lead to a significantly smaller BH mass and accretion rate: e.g., for spin $a=0.97$ and inclination $i=75\degr$ the best-fit masses are $M_{\rm bh} \approx 26 M_\odot$ and $M_{\rm bh} \approx 12 M_\odot$, with effective accretion rates $M_{\rm dd} \approx 1.8 \times 10^{18}$\,g\,s$^{-1}$ and $M_{\rm dd} \approx 0.3 \times 10^{18}$\,g\, s$^{-1}$ for $D_{\rm bh}=11.5$\,kpc and $D_{\rm bh}=4.7$\,kpc, respectively. 
The best-fit value of the BH mass is also impacted by the values of spin and inclination. Table~\ref{table:nicernustar_spinincl} shows the mass and accretion rate results obtained with {\texttt{kerrbb}} for a fixed distance $D_{\rm bh} = 11.5$\,kpc, but for different sets of spin and inclination values. The goodness of the fit clearly shows the preference towards high masses.

Spectral Model 3 (SM3): We then replaced {\texttt{kerrbb}} with the {\texttt{slimbh}} model \citep{SadowskiPhD2011}, that accounts for a vertical structure of the disk using the code {\tt{tlusty}} \citep{Hubeny1998}. This model is suited especially for 
high accretion rates and luminosities $L>0.3\,L_{\rm Edd}$ when conditions for the razor-thin standard accretion disk model are not fulfilled and a very large hardening factor is required to fit well the spectra \citep{Straub2011}.  In XSPEC notation this model can be denoted as {\tt mbpo$\times$edge$\times$tbabs$\times$cloudy$\times$(slimbh+nthcomp)}. The best-fit model describes the data with a $\chi^2$/dof = 3494/2933 (see 
Figure~\ref{fig:simult_nicernustar} and Table\,\ref{table:nicernustar}).
The reduced chi-squared value $\chi^2_{\rm red} = \chi^2$/dof $\lesssim 1.2$ provides a very good fit, given that we have not applied any systematics to the data; these would account for uncertainties in the instrument calibration between {NICER} and {NuSTAR} (that can be within a few percent), as well as for possible spectral variability within the individual exposures (since the data acquired by the two missions were not strictly simultaneous).\footnote{While we have restricted the {NICER} data to be within the {NuSTAR} observations, we did not do that vice versa, since the statistics of such restricted {NuSTAR} data would be too low.}
Variable absorption lines are also a contributing factor to the chi-square in the joint spectral fits of {NICER} and {NuSTAR}. However, as it falls outside the main scope of our study, we did not delve into investigating this aspect in detail. It is noteworthy that these lines do not significantly impact the crucial continuum parameters in our analysis.
The fit requires a BH mass $\sim18\,M_{\odot}$ and a spin $a \sim0.7$, parameters which are more in line with expectations \citep{Seifina2014}. The luminosity is $L \approx 0.5\,L_{\rm Edd}$, i.e. already in the range in which the slim disk approximation is more appropriate than the geometrically thin disk one.
The soft X-ray spectrum is dominated by the thermal accretion disk emission. {NuSTAR} data revealed a variable Comptonisation component, which is best visible in a variable tail at very high energies ($\approx 15$--$30$\,keV). The photon index varies in the range $\Gamma \approx 2.8$--$4.8$, while the soft X-ray spectrum varies only a bit. In the $2$--$8$\,keV energy range, the Comptonisation contributes by $2$--$3\%$ and is almost negligible in the soft X-ray analysis.

\begin{figure}[t]
    \centering
    \includegraphics[height=13cm]{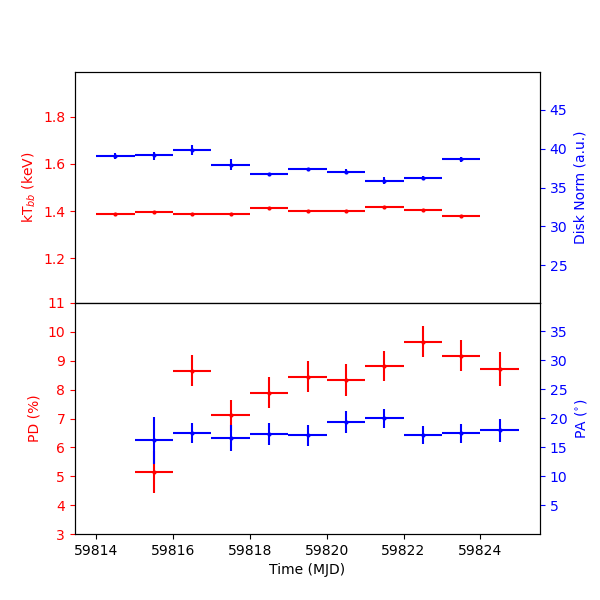}
    \caption{
    Top panel: time variability of the inner disk temperature ($kT_{\rm bb}$, red) and of the BB normalization (blue). Bottom panel: time variability of the polarization degree (red) and of the polarization angle (blue). 
    The error bars are shown at 68\% confidence levels.}
    \label{fig:timevar}
\end{figure}

We also investigated the time evolution of the spectral and polarimetric properties in one-day bins. Here, we characterize the thermal component by its maximum temperature and normalization. Therefore, we base the spectral analysis on fitting the 2--10 keV {NICER} data with the multi-temperature BB disk model \cite[{\sc{ezdiskbb}},][]{Zimmerman_2005}. We use the same \textsc{cloudy} model as in the main spectral fit to model the absorption lines. We neglect the $<2$\,keV data, since absorption strongly suppresses the low-energy flux. The thermal model gives a good fit, as the power law component contributes only by 2--3\% to the 2--10 keV flux. 
The results of the analysis are shown in Figure~\ref{fig:timevar}. 
Whereas the maximum disk temperature stayed rather constant around $\sim 1.4$\,keV, the flux normalization varied between 35 and 40.
The normalization is given by the expression
$f^{-4}\,(R_{\rm in}/D)^2\,\cos{i}$ with
$f$ being the spectral hardening factor,
$R_{\rm in}$ the inner radius of the disk in km, $D$ the source distance in units of 10\,kpc, and $i$ the disk inclination. The polarization degree shows significant variability.
Fitting a constant model to the {IXPE} results gives
a $\chi^2$ of 34.9 for 9 degrees of freedom (chance probability 6$\times 10^{-5}$).


%

\section{Theoretical modeling of the spectro-polarimetric results}\label{sec:theormodels}


Our polarimetric results for 4U 1630-47 are challenging to explain with existing models. Specifically, the PD is very high and increases with energy. In this section, we discuss progressively more complex models for the emergent PD that would be measured by an observer in the rest frame of the radiating plasma.

The PD measured by a distant stationary observer is additionally influenced by general relativistic (GR) effects. There are two main effects. First, as the polarization vector is parallel transported along null geodesics in a curved space-time, the polarization direction projected onto the sky changes, so that the competing contributions coming from different parts of the disk partially cancel each other \citep{StarkConnors1977,Dovciak_2008}. Second, photons following the space-time curvature can return to the disk and scatter off it. Since photons reflected upon the disk are expected to be polarized perpendicularly to those arriving directly to the observer \citep{Schnittman_2009,Taverna2020}, the contributions of direct and returning radiation again partially cancel each other. Both of these effects thus reduce the overall PD from the restframe value, meaning that the models discussed in this section must be able to produce PD in the restframe greater than what we observe for 4U 1630-47 to have a chance of reproducing the data once GR effects are included.

\subsection{Analytical results}
\begin{figure}[t]
    \centering
  \includegraphics[width=0.6\textwidth]{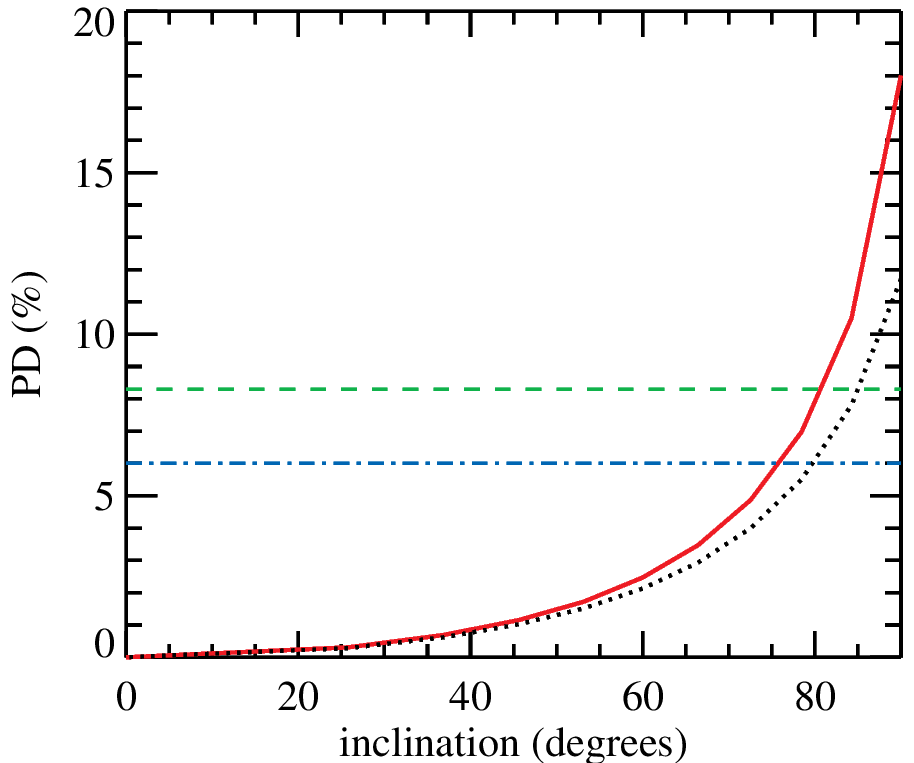}
    \caption{Polarization degree as a function of inclination for the pure electron-scattering atmosphere (black dotted line) compared to the case of partial ionization of matter (absorption effects) and distributed sources (red solid line). The blue dot-dashed and green dashed lines show the most likely values of PD derived in the $2.0$--$2.6$ and $2.0$--$8.0$ keV bands, respectively.}
    \label{fig:es_abs_pol}
\end{figure}

We first consider the classical framework of pure electron scattering in a semi-infinite atmosphere \citep{Chandrasekhar1960,Sobolev1963}. 
In this case, the PD is constant with energy since Thomson scattering is energy independent. 
If the disk is flat and in the absence of relativistic effects the PA is aligned with the disk plane and does not depend on energy. Figure~\ref{fig:es_abs_pol} (black dashed line) shows how PD increases with observer inclination. We see that, even ignoring depolarizing GR effects, the 2--8 keV polarization of 4U 1630-47 (green dashed line) can only be reproduced for an inclination of $\sim 85\degr$. At such high inclinations the X-ray source becomes obscured by the outer parts of the accretion disk, for which an opening angle $\sim12\degr$ has been observed \citep{deJong1996}. Furthermore, complete eclipses are expected in this case, which however have not been detected in this source. Once GR effects are taken into account, the PD is below the observed 2--8 keV values even for an edge-on observer. Therefore, the classical pure electron-scattering atmosphere model fails to reproduce both the energy dependence and overall value of the observed PD.

Additionally considering absorption opacity (i.e. relaxing the over-simplified assumption of complete ionization of matter in the disk atmosphere) can significantly enhance the PD of escaping spectra \citep{LightmanShapiro1975,LoskutovSobolev1979}. The red solid line in Figure~\ref{fig:es_abs_pol} represents the calculation of \citet[][their Table 3, first column]{LoskutovSobolev1981}. Here, true absorption is included analytically by defining the ratio $\lambda$ of the scattering to the total, true absorption $+$ scattering, coefficients and setting it to $\lambda=0.5$. We see that the increase in PD over the pure electron-scattering case is greater for larger inclinations. This is because photons that already entered the disk atmosphere traveling approximately along the inclined observer's line of sight (LOS) before experiencing a scattering with a small scattering angle pass through a greater absorption optical depth on their way through the disk atmosphere, and are therefore more likely to be absorbed, than photons that entered the atmosphere vertically and experienced a scattering high in the atmosphere with a large scattering angle to emerge along the LOS. Absorption therefore increases the relative number of photons that reach the observer via a large scattering angle, thus increasing the PD. Here, the effect of photons being emitted in the atmosphere itself is included. This decreases the PD as compared to the classical Milne problem \citep{Milne1921}, whereby all the photons originate from the bottom of the atmosphere, but we see that the PD is still higher than the pure electron-scattering atmosphere model. The emergent PD is now equal to the observed 2--8 keV PD at an inclination compatible with the maximum inclination expected for non-eclipsing BHXRBs $i_{\rm max}\sim78\degr$ \citep{deJong1996}.

However, the simple absorption model considered here cannot explain the observed PD with a reasonable inclination angle once de-polarizing GR effects are taken into account. Moreover, if we wish to explain an energy dependence of PD, we must account for the energy dependence of the absorption coefficient in place of simply parameterizing the ratio of scattering to total opacity. We must therefore consider a more sophisticated treatment of the disk atmosphere.

\subsection{Radiative transfer calculations of a plane-parallel disk atmosphere embedded in flat space-time}\label{sec:cloudytitan}


We now include energy-dependent absorption in our model, accounting for the effects of partial ionization of different atomic species on the polarization properties within a passive plane-parallel slab on top of a source of unpolarized isotropic single-temperature blackbody radiation. This scenario resembles the Milne problem, but the ionization structure of the medium is pre-computed, rather than parameterised. We derived the one-dimensional vertical ionization profiles for different photo-ionization equilibrium (PIE) regimes using the \verb|TITAN| \citep{Dumont2003} and \verb|CLOUDY| \citep{Ferland2017} codes and for different collisional ionization equilibrium (CIE) regimes using the \verb|CLOUDY| code only, as originally discussed by \citet{Taverna2021} and \citet{Podgorny_2022}. The ionization profile is subsequently used as input for the 3D Monte Carlo, radiative transfer code \verb|STOKES| \citep{Goosmann2007, Marin2012, Marin2015, Marin2018}. We find that the change of the emergent radiation PD with energy strongly depends on the ionization profile itself, rather than on the physical conditions (i.e. either PIE or CIE) which led to a certain ionization profile. We therefore present only results assuming PIE, following \citet{Taverna2021} and \citet{Podgorny_2022}. As the ionization profiles from both codes in PIE agree with each other, we limit the following discussion to the results obtained with the \verb|TITAN| code pre-computations, used in \verb|STOKES|. We adopt the typical solar abundance from \cite{Asplund2005}, with $A_\mathrm{Fe} = 1.0$, which is important for the energy-dependent contribution of ionization edges. We present calculations that assume a constant density throughout the atmosphere. We neglect Compton up-scattering, since it would be effective for higher slab temperatures than those considered in our modeling and the effects would be visible only at energies above {IXPE} band. On the other hand, multiple Compton down-scatterings are included and this process is important in the genesis of polarization within the {IXPE} band.

\begin{figure}[t!]
    \centering
    \includegraphics[height=12cm]{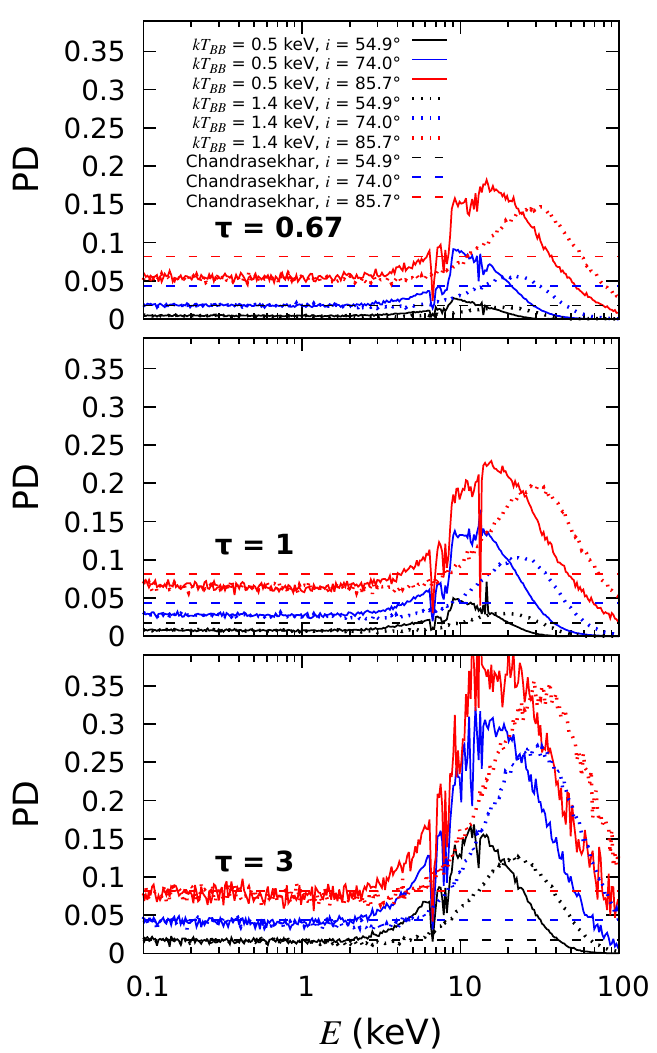}
    \caption{Polarization degree vs. energy from the {\tt TITAN} and {\tt STOKES} 
    codes for a highly ionized slab ($n(\textrm{H}) = 10^{18} \ \textrm{cm}^{-3}$ and a standard BB normalization of the flux at the bottom of the slab). The observer's inclination and optical depth of the slab scale PD according to the energy dependent contribution of both absorption and scattering. The BB temperature of the incident radiation sets weights on different energies. The Chandrasekhar's scattering limit is reached for $\tau \geq 3$ at energies $\leq 2$ keV for all significantly contributing single-color BBs. The obtained PA in all the studied cases is constant with energy and oriented to be parallel to the slab.}
    \label{fig:TS_full_range}
\end{figure}

Figure~\ref{fig:TS_full_range} shows PD of the emergent radiation as function of energy for different slab optical depths $\tau$, observer's inclinations $i$ and temperatures $kT_\mathrm{BB}$ of the illuminating BB, for a highly ionized slab (density $n(\textrm{H}) = 10^{18} \ \textrm{cm}^{-3}$). The optical depth is changed only by changing the height of the layer, ensuring that different optical depths in the plot correspond to the same density and the ionization structure. In this respect, increasing optical depth and inclination show similar effects (both enhancing PD), as photons are on average reaching the observer through larger portions of the slab. We see that the PD increases with energy in the 2--8 keV range, as is observed for 4U 1630-47. The energy dependence is a result of trade-offs between the relative importance of absorption and scattering, and is thus highly dependent on the ionization parameter. Absorption boosts the PD, and so energy ranges where the absorption cross section is larger exhibit a higher PD than the pure scattering limit. From a wider X-ray energy perspective, the cross section for photoelectric absorption declines as $\sim E^{-3}$, while the scattering cross section remains roughly constant until $\approx 50 \ \textrm{keV}$ and should dominate over absorption already above $\approx 0.2 \ \textrm{keV}$. However, in the {IXPE} band this general trend is reversed; around 2 keV the photoelectric absorption is insignificant, in accordance with the general trend, due to the lack of ionization edges, while at higher energies the polarization properties are still strongly affected by absorption (depending on the BB temperature), mostly due to the highly-ionized iron edge at $\sim 9$~keV. We find that this iron edge is still important even for the highest ionization states achievable with the utilized radiative transfer codes under the studied conditions.
 We refer the reader to the detailed study by \citet[see their figures 1 and 4]{Davis2005}, which estimates the energy dependent contributions of various processes to the total opacity in the atmosphere of BHXRB discs. Compton down scattering also contributes to PD increasing with energy, since photons that experience more scatterings have a lower energy and a lower PD.

For the moderate ionizations explored in \cite{Taverna2021}, the PD instead \textit{decreases} in the IXPE bandpass due to high absorption opacity at $\sim 2$ keV (their Figure 3). This $\sim 2$ keV opacity disappears for ionization parameters above $\xi = 4\pi F_\textrm{BB}/n(\textrm{H}) \gtrsim 10^5 \, \textrm{erg\,cm\,s}^{-1}$, leading to the trend that we see in Figure~\ref{fig:TS_full_range} with PD close to the pure-scattering Chandrasekhar-Sobolev limit at $\sim 2$ keV for $\tau \gtrsim 3$. The trend of rising PD with energy that we observe for 4U 1630-47 can therefore only be reproduced for a range of ionization states. The PA is constant with energy and parallel to the slab plane for all parameter combinations trialed. We find that the vertical stratification of the atmosphere in up to 50 plane-parallel layers gives the same results as a single layer, as long as the medium is highly ionized. Additional calculations show that vertically stratified atmospheres with BB temperatures down to $kT_\textrm{BB} = 0.5$ keV, constant densities up to $n(\textrm{H}) = 10^{20} \ \textrm{cm}^{-3}$ and optical thicknesses $\tau\lesssim 10$ give approximately the same net $2$--$8$ keV PD as the simplified approach. The model described above breaks for larger values of the optical depth, since i) the required level of ionization on the distant (not illuminated) side of the passive slab in PIE is no longer reached, and ii) the assumption of a passive slab is not valid due to the lack of internal sources of X-ray radiation inside the slab in our computations. The change in the emergent polarization due to the addition of internal sources is assessed in Appendix \ref{section:radvariation} for a semi-infinite atmosphere leading to the same estimates of locally emergent polarization at 2 keV for inner disc rings.

\begin{figure}[t!]
    \centering
    \includegraphics[height=6cm]{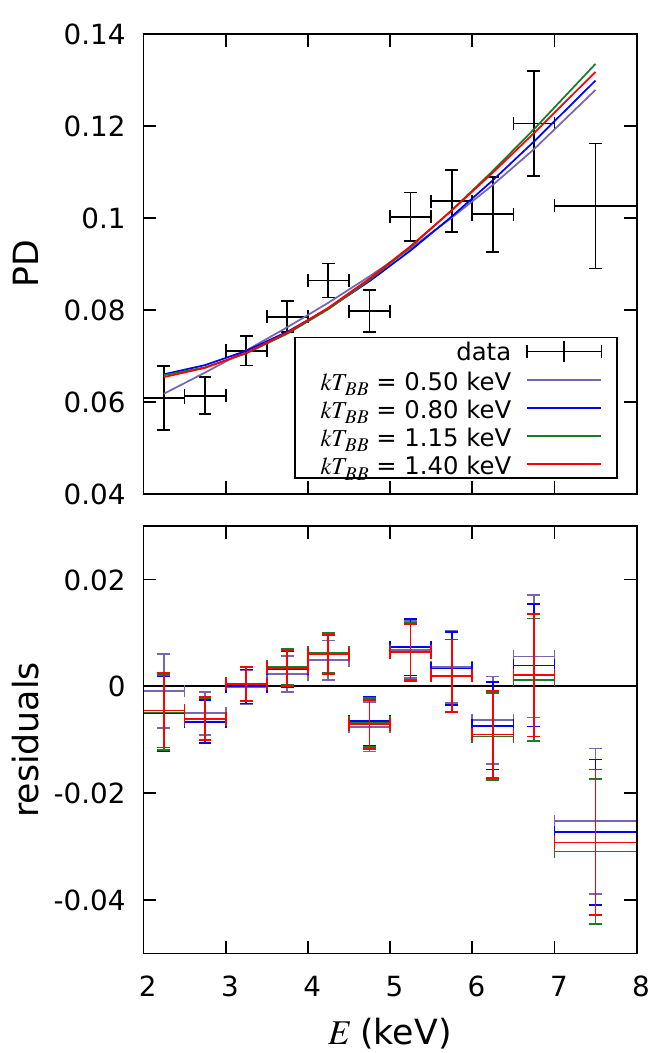}
    \includegraphics[height=6cm]{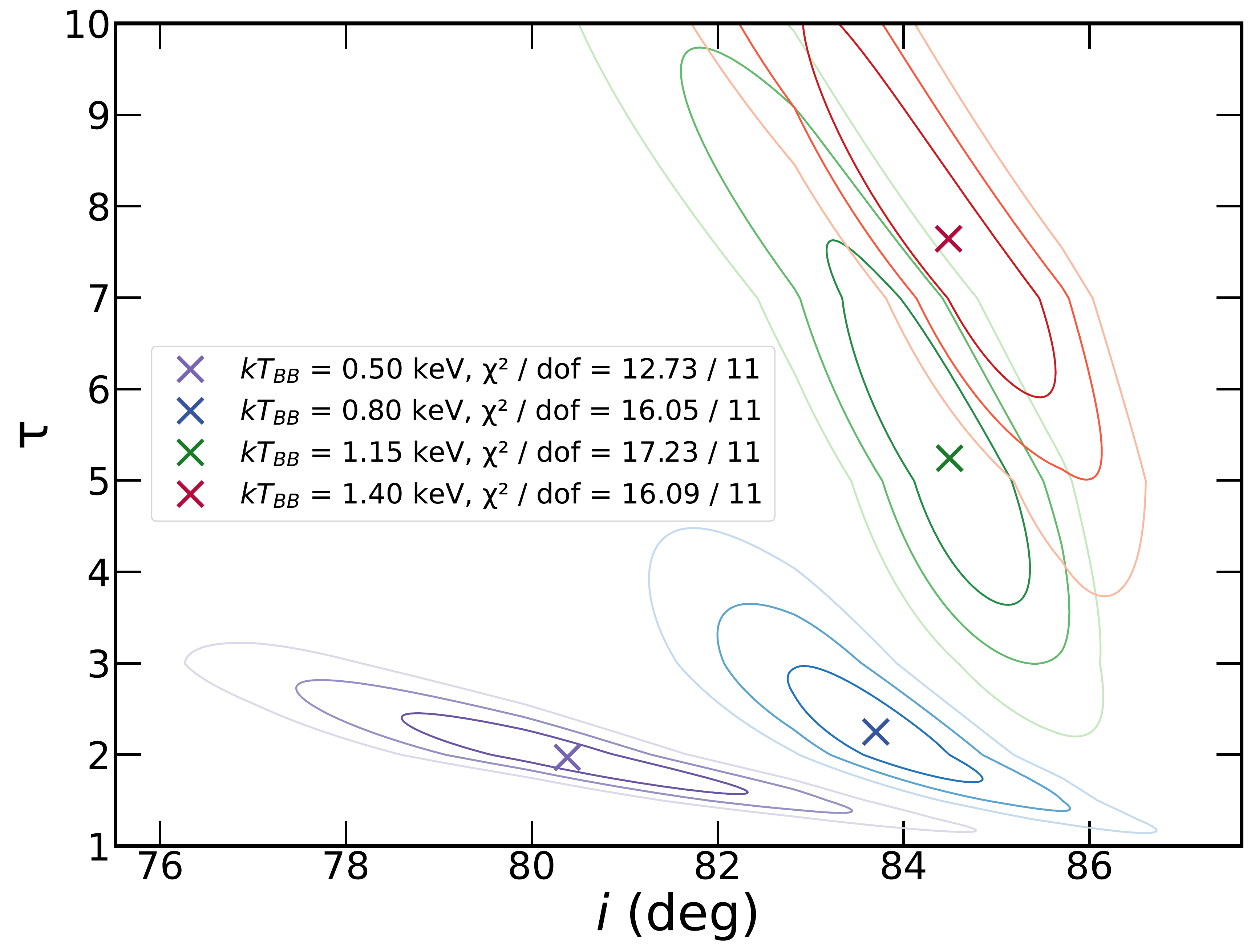}
    \caption{Best-fit local atmosphere solutions of PD vs. energy (left) and 2D contours at $1\sigma$, $2\sigma$, $3\sigma$ levels around their value in the $i$--$\tau$ space (right) obtained with XSPEC by fitting the energy-dependent data with smoothed {\tt TITAN} and {\tt STOKES} models for $kT_\textrm{BB} = 0.50$, $0.80$, $1.15$ and $1.40$ keV.}
    \label{fig:local_best_fits}
\end{figure}

Thus, the observed $6\%$ polarization degree at 2 keV robustly constrains the inclination of the emitting patches to be equal or to exceed $80\degr$ (Figure~\ref{fig:es_abs_pol}). Figure~\ref{fig:local_best_fits} shows a fit of the output from these radiative transfer 
calculations to the {IXPE} data (i.e. ignoring GR effects, approximating the accretion disc
with a plane-parallel slab in flat space. 
The best fits are achieved for optically thick slabs ($\tau \approx 5$) and high inclinations ($i\approx 84\degr$), depending on the temperature of the incident BB radiation. As relativistic effects tend to lower the polarization degree, even higher local optical depths and/or inclinations are needed in case of relativistic accretion discs. 

\subsection{Outflowing disk atmosphere} \label{sec:outflowingdiscatmosphere}
To enhance the degree of local polarization in these models for achievable inclination angles, it would be necessary to hypothesize a greater emission angle in the local reference frame co-moving with the accretion disc. This could occur in case of an outflow with relativistic speed in the vertical direction, i.e. perpendicular to the disc. A similar analogy to this could be the out-flowing corona models at relativistic velocities proposed to explain the {IXPE} observations of Cygnus X-1 \citep{Poutanen_2023}. Due to the relativistic aberration effect, the photons must be emitted with higher emission angles to reach the same observer at a given inclination angle. Thus, we assume a disk atmosphere with decreasing radial profile of the vertical outflow velocity, i.e. $\beta(r)=\beta_0\, r^{-q}$ with $\beta$ being the speed in units of $c$ in the vacuum. This will add two additional parameters, $\beta_0$ and $q$, influencing the predicted polarization properties. The inclusion of this relativistic aberration to the two models and a comparison to the data are shown in the next section. 

\section{Comparison of our models with the data} \label{sec:comparisonmodeldata}


We now include all GR effects (see Appendix \ref{section:radvariation} for details) and compare our model predictions with the observed data.
Since system properties such as inclination, distance, BH mass and spin are not measured for this object, but rather estimated indirectly, we study the polarization properties of our models assuming several values of high inclination and BH spin. Once the rest of the parameters that can change the spectral shape are derived from the fit of the observed spectrum, we fit the energy-dependent PD and PA, checking if any constraint can be put {\em a posteriori} on the spin and/or on the inclination. Since in this case we are interested in the thermal spectral component, and given that {NuSTAR} data only cover a small fraction of the {IXPE} observation (that we used to estimate the small contribution of the non-thermal component to the total flux), the spectral fits analysed in this section refer to the {NICER} data set only. We emphasise that, excluding the non-thermal component, in the current analysis we use all the (summed) {NICER} data, i.e. not only the ones corresponding to the {NuSTAR} observations as in Section \ref{section:specdataanalysis}. Moreover, since the spectra observed by {NICER} are much better calibrated than those observed by {IXPE} (especially for high flux sources), we use the {IXPE} data only for polarization. 
We fit the {NICER} spectra using the model {\sc edge$\times$tbabs$\times$cloudy(kynbbrr)}. Here, we use the Novikov-Thorne geometrically thin disk model {\sc kynbbrr} \cite[][instead of and similar to the {\sc kerrbb} model]{Dovciak_2008, Taverna2020, Mikusincova2023} for the thermal component, since this model allows us to study also the polarization properties in different scenarios. The other model components are the same as already described in section \ref{section:specdataanalysis}. 
In the current spectral analysis we use $1\%$ systematic error for the {NICER} data. Note that the spectral fit does not require the Comptonized component at these energies, confirming that its average contribution is very low in the full {IXPE} observation. The best spectral fits are summarised in the Table~\ref{table:nicer_spinincl}.

\begin{figure}[thb]
    \parbox[b]{0.5\textwidth}{\centering
    \includegraphics[width=0.43\textwidth,trim={1.2cm 5.2cm 0.9cm 0},clip]{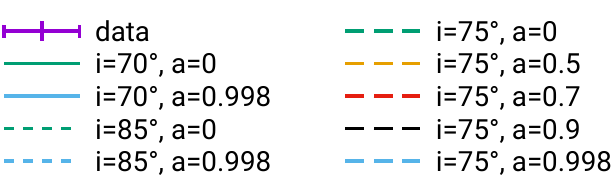}\\[0.6cm]
    \includegraphics[width=0.51\textwidth,trim={0 0 0 1.9cm},clip]{kynbbrr-NTCh-A-PD.pdf}}
    \parbox[b]{0.5\textwidth}{\includegraphics[width=0.51\textwidth,trim={0 0 0 1.9cm},clip]{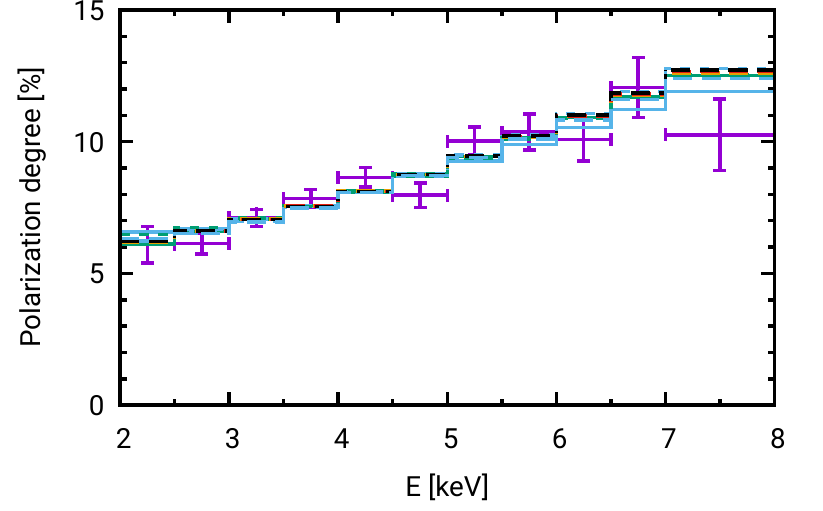}\\[1mm]
    \includegraphics[width=0.51\textwidth,trim={0 0 0 1.9cm},clip]{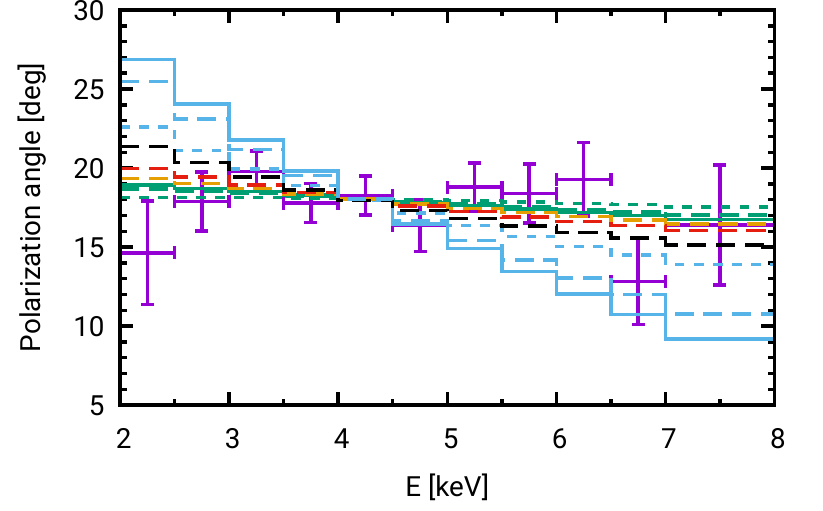}}
    \caption{Comparison of the observed (purple points error bars at 68\% confidence level) and modeled 
    4U\,1630--47 polarization properties (different lines).
    The standard Novikov-Thorne geometrically thin accretion disk model, with emission and reflection according to Chandrasekhar's pure scattering atmosphere model, significantly under-predict the PD (left). The results of thin disk models with an outflowing, partially-ionized atmosphere can explain the data
    for low and intermediate spins ($a\,=\,$ 0, 0.5 or 0.7) and $i\,=\,70\degr\,$ or $75\degr$ inclinations (top-right and bottom-right panels).
    The models shown in the right panels assume optical thickness $\tau = 7$ and an outflow velocity $v\sim0.5\,c$.}
    \label{fig:model1}
\end{figure}

In the next step we froze all the spectral parameters which influence the spectral shape and tried to fit the observed PD and PA. 
To this purpose, we have developed different flavors of the {\sc kynbbrr} polarization model that originally assumed Chandrasekhar approximation of pure scattering atmosphere for direct radiation and Chandrasekhar multi-scattering approximation for polarization of reflected disk self-irradiation with an assumed albedo of 0.5. We denote this original model as model (A). The orientation of the system on the sky of the observer is the only free parameter of this model to fit the observed polarization properties. It defines the direction of the system rotation axis.
Figure~\ref{fig:model1} (left panel) shows the results of fitting model (A) to the observed energy-dependent PD (parameters in Table~\ref{table:ixpe_chi2}). We see that (as mentioned in section \ref{sec:theormodels}),
the PD predicted by this model is too low (see the left panel on Figure~\ref{fig:model1} and Table~\ref{table:ixpe_chi2}).

To increase the local PD, we enhance the model (A) by assuming a vertical outflow with relativistic speed as we described in detail in section \ref{sec:outflowingdiscatmosphere}. The predicted polarization properties of this model, denoted as (B), are shown in Figure \ref{fig:model2} and Table \ref{table:ixpe_chi2}. This model still does not fit the observed polarization in a satisfactory way, failing to reproduce the PD increase with energy.

The third flavor of the polarization model, denoted as (C), uses the local polarization degree derived by the model described in section \ref{sec:cloudytitan}, i.e. an ionized passive slab with finite optical depth and constant density computed with the \verb|TITAN| and \verb|STOKES| codes. The local PD in this model depends on the disk temperature (given by the Novikov-Thorne temperature profile), emission angle (computed through a ray-tracing technique in the curved space-time for a given observer inclination) and optical thickness of this layer. The latter is a new parameter that influences the predicted polarization properties. As mentioned in the previous sections, also this model does not fit the observed polarization properties (see the right panel on Figure~\ref{fig:model2} and Table~\ref{table:ixpe_chi2}): although the increase of PD with energy corresponds to what is observed, the PD magnitude is too low. 
In order to fix this issue, we increased the local PD by assuming vertical outflow velocity that causes higher local emission angle due to aberration effects, as in the model (B); we denote this as a model (D). The top-right panel on Figure~\ref{fig:model1} and Table~\ref{table:ixpe_chi2} show that this model can fit the observed data quite well, except for highly spinning BHs, for which the fit of the observed PA is not satisfactory (see bottom-right panel of Figure~\ref{fig:pol_angle}). Note that the PA energy dependence predicted with different flavors of the {\sc kynbbrr} model is very similar due to the fact that the local polarization angle is always assumed to be parallel with the disc. The predicted PA is thus dependent mainly on the relativistic effects. Acceptable fits are denoted in bold-face in Table \ref{table:ixpe_chi2} and the best fit parameters are shown in Table \ref{table:ixpe_params}. Note that, in order to reproduce the observed PD, the scattering layer has to be outflowing with approximately half of the speed of light.

\section{Discussion and conclusions}\label{discussion}
The high, energy-dependent PD of 4U\,1630--47 cannot be explained in the framework of the standard thin disk model, where the polarization of the thermal disk emission should roughly follow Chandrasekhar’s classical result for a semi-infinite, free electron scattering atmosphere \citep{Chandrasekhar1960}. Instead, radiative transfer calculations of a plane-parallel, partially-ionized slab in photo-ionization equilibrium \citep{Taverna2021, Podgorny_2022}, extensively discussed in the section \ref{sec:cloudytitan}, can predict a PD in the disk rest-frame increasing from $\sim 6\%$ at 2\,keV to $\sim 10\%$ at 8\,keV (as observed), as long as the slab is highly ionized and viewed from an angle of $\sim 84\degr$. The former  condition is met for $k\,T_\textrm{BB} \gtrsim 0.5$ keV, low ($\lesssim 10$) optical depths and for a wide range of slab hydrogen densities, $n(\textrm{H})\sim 10^{12}$--$10^{20} \ \textrm{cm}^{-3}$. In this scenario, the polarization angle should be parallel with the disc, although this cannot be assessed as yet, due to the lack of constraints on the orientation of the radio jet. However, GR effects reduce the predicted observed PD in the {IXPE} energy band to well below the observed values (Figure~\ref{fig:disc_atm}, and Figure~\ref{fig:model2}, right panel).



Additional corrections must be considered, alongside those already discussed.
We thus further assume particle outflows with relativistic vertical motion from the atmosphere of the disc. 
Due to relativistic beaming effects, this model will predict a higher PD. 
Including this into the partially-ionized slab model, we can explain the observed PD and PA (right panels of Figure\,\ref{fig:model1})  with the combination of a low or intermediate BH spin ($a\lesssim0.7$), and a highly-ionized atmosphere with an optical depth of $\tau\sim 7$, outflowing perpendicular to the accretion disk at a velocity $v\sim0.5\,c$. 
Note that a passive disc atmosphere is considered here with no internal sources of radiation, although the disc is quite hot.
In Appendix \ref{section:radvariation}, we consider internal sources of radiation within the semi-infinite disc atmosphere in a different model, but without any outflowing velocities, which also provides too low PD to explain the observed values. More detailed modelling is deferred to a future work.

The observed PD variability (see Figure~\ref{fig:timevar}) could then be explained by changes in the optical depth or due to varying outflow velocity of such an atmosphere. Consequently, the optical depth and the outflow velocity are anti-correlated. Essentially, an increase in optical depth is analogous to a high inclination angle, just as increasing the outflow velocity would be. Hence inclination angle, optical depth and outflow velocity are all anti-correlated to each other.

The absorption lines imprinted on the spectrum with blue shift velocity $v \sim 0.003~c$ originate from an equatorial wind located farther from the inner disk than the rapidly outflowing atmosphere. Previous analysis of the absorption lines using various instruments has indicated that absorption lines (especially Fe XXV and Fe XXVI) are found at a distance of $10^3$--$10^4$ gravitational radii ($R_{\rm g}$) and with a mass outflow rate of 40\% of the accretion rate \citep{King_2014,Trigo_2014,Miller_2015}. Later, \cite{Fukumura_2021} modeled these lines using a Magneto-hydrodynamic (MHD) model, which starts at ISCO and is launched across the disc domain. Within that model, the inner regions of the outflow exhibit high relativistic velocities, which gradually decrease as a function of radial distances ($V\propto r^{-0.5}$). The matter is highly ionized close to the black hole, and the absorption lines are found at a radius of $10^5R_{\rm g}$, where the ionization drops. Moreover, the velocity observed in the absorption lines represents the velocity along the line of sight (LOS). However, the outflow may possess a higher velocity component perpendicular to the accretion disk, consequently resulting in a smaller velocity into high inclination angles, as in the case of this source. It is plausible that the highly ionized and outflowing disk atmosphere could be connected to the absorption lines found at larger distances i.e. the outflow is initially rapid and dense at the disk surface before spreading and slowing down at large distances like in the case of magneto-hydrodynamic (MHD) winds \citep{Blanford1982,Contopoulos1994,Ferreira_1997,Fukumura_2010, Chakravorty_2016}. In a very approximate estimation, assuming an outflow with a vertical velocity (perpendicular to the disc plane) of $0.5c$ (as in the case of the outflowing disc atmosphere) at a distance of $1R_{\rm g}$, it should result in a LOS velocity of $0.13 c$ at an inclination angle of 75\degr. If we assume a radial dependence of $V\propto r^{-0.5}$ as in MHD models, this velocity will decrease to $0.004 c$ at a distance of $10^3R_{\rm g}$, consistent with the velocity of outer outflow found in this source.  However, there has been no observational evidence indicating a connection between the inner and outer outflows. Conducting additional observations at varying continuum and line flux levels during different states would be necessary to investigate potential correlations between them.

Another possible way to explain the high polarization and its energy dependence would be increasing the contribution of the polarized flux through reflection due to disk self-irradiation. The underlying principle here is that the harder X-rays coming closer to the black hole would have higher fraction of photons returning back to the disc and getting reflected and hence higher polarization. Since this is dependent on the amount of self-irradiation, this can be larger only if we change the disk geometry, specifically if the disk scale-height is substantially larger than in the geometrically thin accretion disk (as seen as well in Appendix \ref{section:slimdisc}). The modeling described in Appendix \ref{section:slimdisc} shows that geometrically thicker discs, such as the slim disk \citep{Abramowicz_1988} or recently very popular puffy discs \citep{Wielgus_2022}, can lead to a large PD, increasing with energy. However, for the considered disk geometries, even our most optimistic fine tuning of the model parameters (BH spin, inclination of the observer, disk scale height) cannot fully reproduce the observed high PD. We cannot exclude here that the high PD may be explained with other disk geometries, that can increase the fraction of scattered X-rays even more.

\begin{figure}[thb]
    \parbox[b]{0.5\textwidth}{\centering
    \includegraphics[width=0.43\textwidth,trim={1.2cm 5.2cm 0.9cm 0},clip]{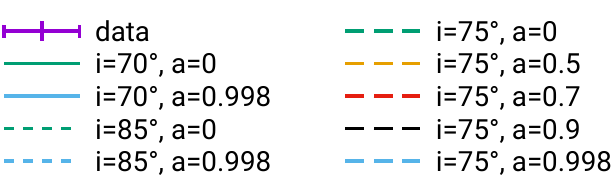}\\[0.6cm]
    \includegraphics[width=0.51\textwidth,trim={0 0 0 1.9cm},clip]{kynbbrr-NTCh-A100-PD.pdf}}
    \parbox[b]{0.5\textwidth}{\includegraphics[width=0.51\textwidth,trim={0 0 0 1.9cm},clip]{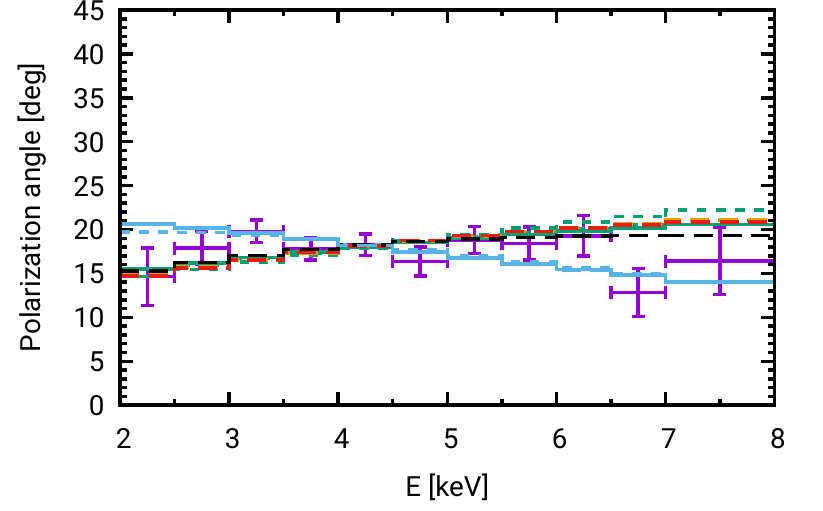}\\[1mm]
    \includegraphics[width=0.51\textwidth,trim={0 0 0 1.9cm},clip]{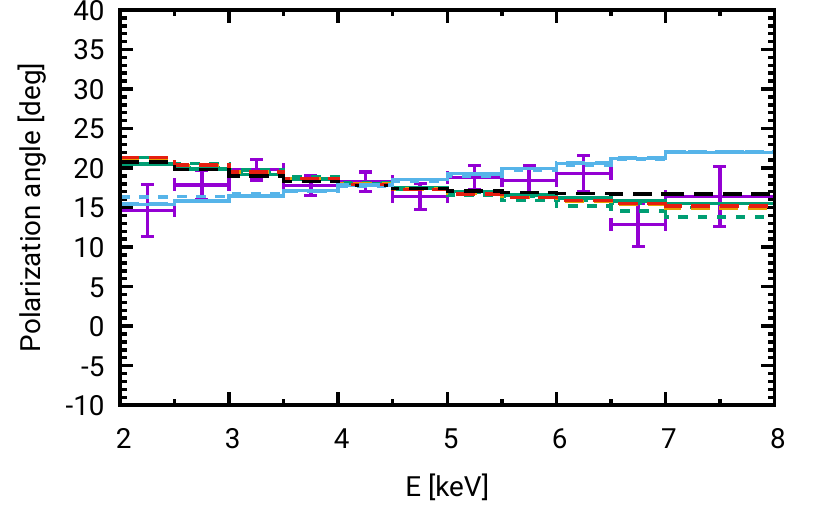}}
    \caption{Energy-dependent polarization degree (left) and polarization angle for system rotation anticlockwise (top-right) and clock-wise (bottom-right) in the case of a geometrically thin disk with the contribution of the reflected self-irradiation, increased by allowing the albedo to be larger than 1. 
}
    \label{fig:model3}
\end{figure}

To provide a rough estimate of the required increase in self-irradiation, we can use the geometrically thin disk scenario where the albedo will be phenomenologically allowed to be larger than one. The results of this experiment are shown in Figure~\ref{fig:model3}. While the exact shape of the model predicted polarization degree and angle and thus also the dependence of the fit statistics on model parameters should be taken with caution, since the model is not self-consistent any more, we can still discuss the value of the albedo and its interpretation with respect to amount of self-irradiation. The required excess in albedo is the lowest for the highest studied spin value and increases for lower spin values while it depends much less on the inclination. This is due to the fact that the amount of self-irradiation is already much larger in case of the disk reaching closer to the black hole horizon for highly spinning BH thus smaller multiplicative factor (albedo) is needed to reach the needed increase in polarized reflected flux. On the other hand the amount of the reflected self-irradiation depends much less on the inclination. For the inclination of $75\degr$, the fitted albedo value was 4.5, 11.4, 17.1, 22.2 and 36.2 for the spin values of 0.998, 0.9, 0.7, 0.5 and 0, respectively. In the case of highly ionized accretion discs, the albedo is expected to be in the range of 0.5 to 1. Hence a flux increase of the reflected self-irradiaton by a factor of at least 5 to 10 (in case of very highly spinning BH) with respect to the highly ionized geometrically thin disc would be required to explain the observed polarization. On the other hand, partial obscuration of the inner disc by the disc itself (self-shadowing) would decrease the contribution of the direct emission and in this way increase the relative contribution of the reflected self-irradiation component.

The high inclination could also be explained by a scenario where the inner accretion flow is more inclined than the binary system, like in the case of a wrapped disc. If the BH spin axis is not aligned with the binary orbital axis, the gravitational effects can result in the inner accretion flow becoming perpendicular to the BH spin axis, potentially leading to a greater inclination of the inner disk \citep{Bardeen_1975}.

Our spectral analyses yielded a notably unphysical photon index, particularly in the third {\it NuSTAR} observation. Nevertheless, some studies have proposed a non-thermal/hybrid Comptonization model to explain the power-law tail observed in the soft state of BHBs, with a substantial contribution from the Comptonization from the non-thermal electron distribution \citep{Gierlinski_1999,Zdziarski_2001}. 
Although the photon index values from our spectral analysis 
are phenomenological, 
they serve well 
our aim to understand the contribution from power-law in the IXPE energy band of 2--8 keV. Since the basic parameters of this system are not well known, especially the orientation of the system, it is difficult to understand the location of the corona. In case of the hard state of the BHB Cyg X-1, the polarization angle aligned with the jet indicated that the corona was extended over the disk \citep{Krawczynski_2022}. However, the geometry and the location of corona in the soft state is not understood well. Another observation of this source in the Steep power law (SPL) state dominated by the Comptonization component indicates that the Comptonization component is polarized in the same direction (perpendicular to the symmetric axis) and hence can be vertically extended along the symmetric axis of the system \citep{Rodriguez_2023}. Indeed, the disk atmosphere can transition to a state dominated by inverse Compton scattering as the electrons gain energy, which can account for this behavior \citep{Rodriguez_2023}. We refer the reader to the following paper on the SPL state of the source for a discussion on the geometry of the corona \citep{Rodriguez_2023}.

Two other works have independently analysed the same {IXPE} data sets \citep{Kushwaha2023,Rawat2023}, obtaining observational results consistent with those presented in this work. The authors posit that the high polarization degree can be explained by scattering off a wind. However, reflection off a highly ionized wind via Thomson scatterings leads to rather constant PDs
\citep{Ratheesh_2021,Veledina_2023}, contrary to what is observed. Further, for the reflected flux to make a significant contribution to the total spectrum, the solid angle subtended by the wind to the X-ray source needs to be large; in this case, emission lines should be present,   
which we do not see in the {NICER} and {NuSTAR} energy spectra. Recently, \cite{Tomaru_2023} studied a possible wind contribution to the polarization and conclude that the wind cannot explain very high polarization degrees observed in this source. Finally, in a subsequent observation of the source, when it exhibited a steep power law state and no evidence for a wind was apparent, the polarization degree and its energy dependence were similar to the high soft state \citep{Rodriguez_2023}, again arguing against a wind origin for the polarization.
Additionally, we also confirm the negligible contribution of the Comptonized power-law in 2--8 keV band using broad energy-band data of {NuSTAR}.

We conclude that the new observational diagnostics provided by {IXPE} 
imply significant deviations from the standard thin disk model. 
Additional observations of the source in the same emission state at different flux levels and in different emission states will constrain the accretion disk geometry and properties even further. 


\section*{Acknowledgements}
We thank the anonymous reviewer for the suggestions in improving this manuscript. 
The Imaging X-ray Polarimetry Explorer (IXPE) is a joint US and Italian mission.  The US contribution is supported by the National Aeronautics and Space Administration (NASA) and led and managed by its Marshall Space Flight Center (MSFC), with industry partner Ball Aerospace (contract NNM15AA18C).  The Italian contribution is supported by the Italian Space Agency (Agenzia Spaziale Italiana, ASI) through contract ASI-OHBI-2022-13-I.0, agreements ASI-INAF-2022-19-HH.0 and ASI-INFN-2017.13-H0, and its Space Science Data Center (SSDC) with agreements ASI-INAF-2022-14-HH.0 and ASI-INFN 2021-43-HH.0, and by the Istituto Nazionale di Astrofisica (INAF) and the Istituto Nazionale di Fisica Nucleare (INFN) in Italy.  This research used data products provided by the IXPE Team (MSFC, SSDC, INAF, and INFN) and distributed with additional software tools by the High-Energy Astrophysics Science Archive Research Center (HEASARC), at NASA Goddard Space Flight Center (GSFC). 

A.R. and F.T. thank Prof. Keigo Fukumura for useful discussions. M.D., J.Pod., J.S. and V.K. acknowledge the support from the Czech Science Foundation project GACR 21-06825X and the institutional support  from the Astronomical Institute of the Czech Academy of Sciences RVO:67985815. J.Pod. and V.K. thank the Czech-Polish mobility program (M\v{S}MT 8J20PL037 and PPN/BCZ/2019/1/00069) and the European Space Agency PRODEX project 4000132152. M.D. and J.Pod. specially thank Prof. Agata R\'o\.za\'nska for explanations and guidance using the \verb|TITAN| code.
H.K., N.R.C. and A.W. acknowledge NASA support through the grants NNX16AC42G, 80NSSC20K0329, 80NSSC20K0540, NAS8-03060, 80NSSC21K1817, 80NSSC22K1291, and 80NSSC22K1883 as well as support from the McDonnell Center for the Space Sciences. 
A.V., J.Pou., V.L. and S.S.T. acknowledge support from the Academy of Finland grants 349144 and 355672.  
V.F.S. acknowledge Deutsche Forschungsgemeinschaft (DFG) support through the  grant WE 1312/59-1 and the German Academic Exchange Service (DAAD) support through the project 57525212. 
A.I. acknowledges support from the Royal Society. The French collaborators acknowledge financial support from the French High Energy national program from CNRS and from the French Space Agency (CNES).
M.N. acknowledges the support by NASA under award number 80GSFC21M0002.

{\bf Data availability:} The {IXPE}, {NICER} and {NuSTAR} data used in this analysis are from the HEASARC data archive. 
{\bf Code availability:} The codes used to reduce and analyze the data is readily accessible on the HEASOFT web page (\url{https://heasarc.gsfc.nasa.gov/docs/software/lheasoft/download.html}) and within the {\it ixpeobssim} repository (\url{https://ixpeobssim.readthedocs.io/en/latest/?badge=latest.494}). Kindly address any correspondence and requests concerning the remaining codes to  \href{ajay.ratheesh@inaf.it}{ajay.ratheesh@inaf.it}.

\appendix

\renewcommand{\thefigure}{A\arabic{figure}}
\renewcommand{\thetable}{A\arabic{table}}
\setcounter{figure}{0}
\setcounter{table}{0}

\section{Data sets and data reduction}\label{section:datareduction} 

\subsection{IXPE}
 
The Imaging X-ray Polarimetry Explorer ({IXPE}) consists of three independent telescopes, each made up of a Mirror Module Assembly (MMA) and a detector unit (DU) \citep{Ramsey_2019,Soffitta_2021, Baldini_2021,Weisskopf_2022}. {IXPE} observed 4U 1630--47, from 2022 August 23 23:14 to 2022 September 02 18:54, for a total effective exposure of approximately 463 ks. The analysis of the {IXPE} data is performed using the \verb|ixpeobssim| software, version 28.4.0 \citep{Baldini2022}, which is based on level-2 processed data. For our analysis, we used the combined data sample collected by the three identical DUs, with appropriate rotation to align them with the same reference system in sky coordinates. We utilised the SAOImage DS9 software \citep{ds9} for the source and background region selection process. The source region was chosen as a circular area with a radius of $1.0\arcmin$, centred at the region of maximum intensity within the field of view, consistent with the source location. The background region was defined as a concentric annulus with inner radius of $2.5\arcmin$ and outer radius of $4.3\arcmin$. The \verb|ixpeobssim| routine \verb|xpselect| was used to produce the source and background event files.

The polarization degree and angle
were computed using the \verb|xpbin| routine of \verb|ixpeobssim|, using the flag {\tt --algorithm PCUBE}. Version 11 of the {IXPE} response functions was used to process the data. This approach enabled the calculation of the polarization properties in a model-independent manner. The \verb|xpbin| routine allows us to perform the background subtraction from source files: note that the source-emission leakage into the background region does not affect the final source polarization estimation significantly. 
Then we have generated the OGIP standard FITS files of polarization degree and angle together with the unit response files with the \verb|flx2xsp| tool from the \verb|HEASOFT| package \citep{Heasarc_2014} so that we can directly feed them to \verb|XSPEC| \citep{Arnaud_1996} in order to perform polarimetric fits.
The \verb|xpselect| and \verb|xpbin| routines were also used to generate event files for the analysis of variability over time.\\


\subsection{NICER}

{NICER} uses 52 silicon drift detectors (SDDs), each one with a paired single-scattering concentrator optic and mutually aligned on the sky \citep{nicer}; it is sensitive in the 0.2--12 keV range, offering $<100$ ns time resolution, and has a peak effective area $\sim 2000$~cm$^2$. {NICER} carried out 11 observations of 4U~1630--47 during the {IXPE} campaign, from 2022 August 22 to 2022 September 1. A total of 64 good time intervals (GTIs) have been used for our science analysis, for an aggregate time of $\approx27$~ks and a total of $>11$ million X-ray counts.

{NICER} data were reduced and processed with the version 9 of the data analysis software \verb|NICERDAS|. Data were filtered following standard practices, but allowing data from South Atlantic Anomaly (SAA) passages. For each GTI, data from detectors 14, 34, and 54 were excised owing to calibration problems among these subsets. Additionally, the average rates of overshoot, undershoot and X-ray events per GTI were assessed, and any detector which $>15$ median absolute deviation (MAD) was excluded for that GTI. All the exposure times were corrected for the detector dead time ($<1\%$). The background spectra were computed using the 3C~50 background model \citep{Remillard_3C50}.  Only GTIs of length $t>60$ s and for which the background rate was 100 times below the source rate were used for the analysis. A total of 27~ks of simultaneous {NICER} observations were finally available for the analysis. The spectral and light curve files were extracted from the event files using \verb|XSELECT| version 2.5b and the response files were generated using \verb|nicerarf| and \verb|nicerrmf|.\\ 

\subsection{NUSTAR}

4U 1630--47 was observed three times  by {NuSTAR} during the {IXPE} pointing (on 2022 August 25 and 29, and on 2022 September 1) using the two co-aligned X-ray telescopes, each one with a corresponding Focal Plane Module A (FPMA) and B \cite[FPMB,][]{Nustar_2013}. The total elapsed times of the three snapshots are 38.3 ks, 31.6 ks and  32.5 ks, respectively. The Level 1 data products were processed with the Data Analysis Software \verb|NuSTARDAS| package (v. 2.1.2). Cleaned event files (level 2 data products) were produced and calibrated using standard filtering criteria with the \verb|nupipeline| task and the {NuSTAR} calibration files {\sc 20220510}, available in the CALDB database. Extraction radii for the source and background spectra were $60\arcsec$. The spectra and light curves were then generated by the \verb|nuproducts| routine of the \verb|NuSTARDAS| package. FPMA and FPMB spectra were binned following the procedure described in \cite{kb16}, in order to have a signal to noise ratio (SNR) greater than 3 in each spectral channel. The net observing times for the FPMA/FPMB data sets are 16.3 ks/16.6 ks, 13.2 ks/13.4 ks and 14.8 ks/15.0 ks for the three sets of {NuSTAR} data, respectively.

\section{Modeling the effects of the impact of radial variations of the properties of the disk atmosphere, and relativistic effects}\label{section:radvariation}

\begin{figure}
    \centering
    \includegraphics[width=0.65\textwidth]{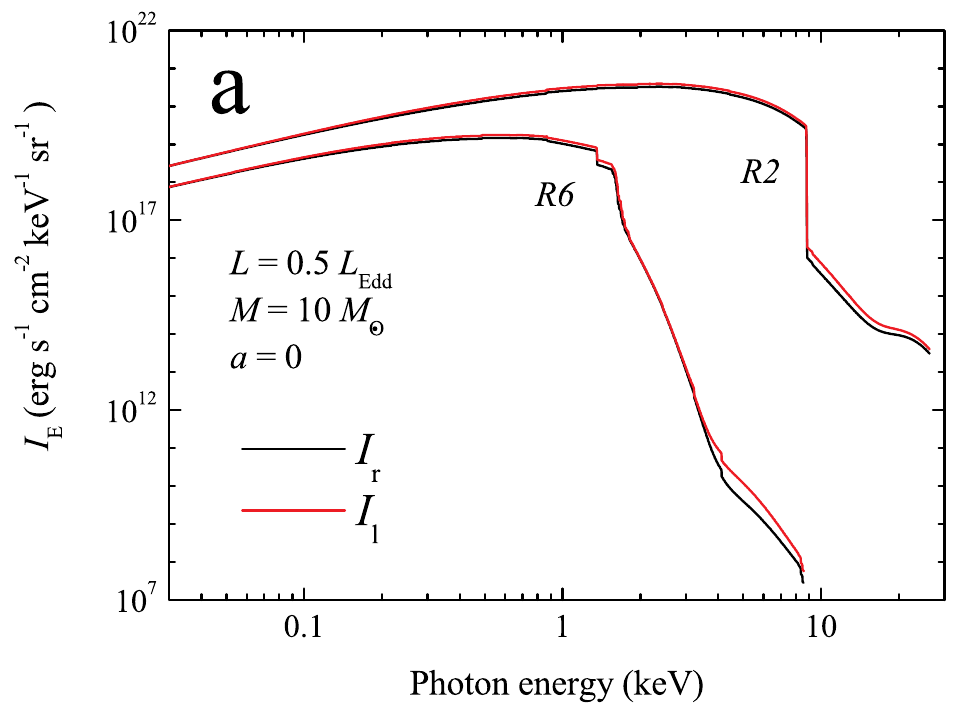}
    \includegraphics[width=0.45\textwidth]{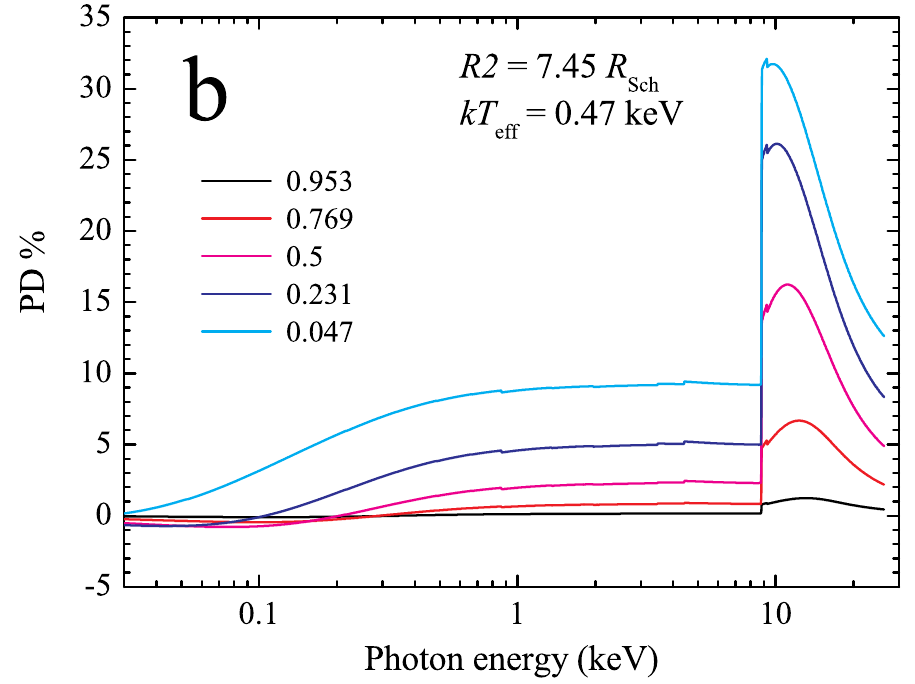}
    \includegraphics[width=0.45\textwidth]{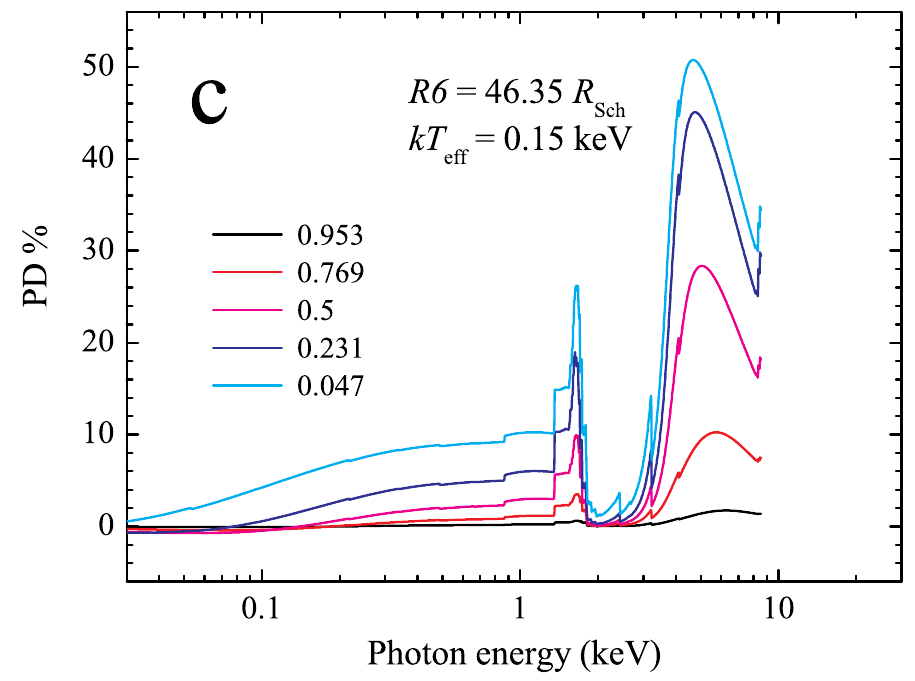}
    \caption{
    Spectra of two rings of the considered disk model (see text) in two modes (a), and PD for the same rings (b and c). The results in panels (b) and (c) are shown for five angles to the local normal; the cosines of the corresponding angles are reported in the panels. 
    }
    \label{fig:ring_pol}
\end{figure}

\begin{figure}
    \centering  \includegraphics[width=0.55\textwidth]{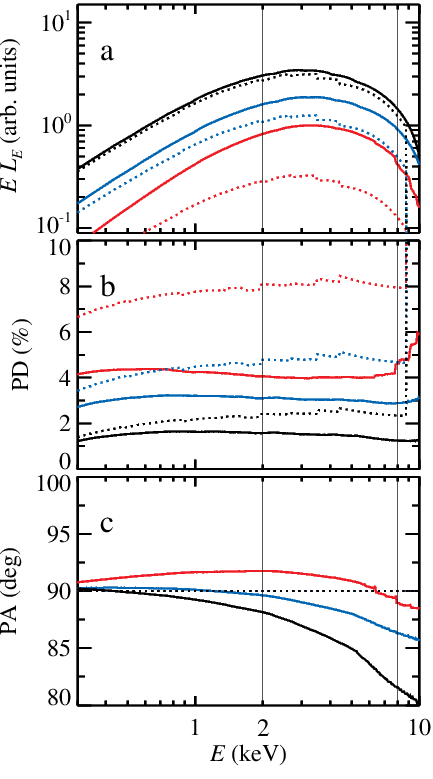}
    \caption{Spectra, PD and PA from a self-consistent model
    of disk atmosphere in radial bins.
    Dotted and solid lines correspond to quantities in the local reference frame and modified by the relativistic effects (as seen by a distant observer), respectively. 
    The results are shown for three disk inclinations: $i=60\degr$ (black), $75\degr$ (blue) and $85\degr$ (red).}
    \label{fig:disc_atm}
\end{figure}

In this section we calculate the polarization of the emission from accretion discs with  realistic radially structured disk atmospheres accounting for relativistic effects \citep{Suleimanov2002,Suleimanov2007}.
In these calculations, the radial disk structure follows from the conservation laws of mass, angular momentum and energy in accordance with the standard Shakura-Sunyaev $\alpha$-disk model \citep{Shakura_1973} and the relativistic corrections by \citet[see also \citealt{Riffert_95}]{Novikov_1973}. The disk model was divided into a number of rings and the vertical structure of each ring was calculated in the grey approximation, using the model atmosphere approach \citep{Suleimanov_92}. The model input parameters were taken from the disk parameters at a given radius: effective temperature, half-thickness, and surface density. The ring was assumed to be in hydrostatic and radiative equilibrium. It is also assumed that the local vertical energy release rate
is proportional to the local pressure. Then we solve the radiation transfer equation in the $0.1$--$20$ keV band (divided into 500 frequency points) using the grey ring model. We adopted the two-mode approximation, with electron scattering 
being the only source of polarization \cite[see details in][]{Suleimanov_23}, and assumed no illumination from the top of the atmosphere, imposing a (mirror) reflection boundary condition in the mid-plane of the disc.
The plasma equation of state is used assuming local thermodynamic equilibrium. We account for the 15 most abundant chemical elements assuming the solar chemical composition  \citep{Asplund_09}; the ionization state of the elements and the excited level populations were computed using the Saha-Boltzmann equations, including pressure ionization effects \citep{Hubeny_94}. 
The absorption 
opacities were calculated accounting for 
bound-free and free-free transitions of all the elements. 
In Figure~\ref{fig:ring_pol}, we present the locally emitted energy spectra for 
two exemplary rings and
five inclinations in the case of
$M_{\rm bh}=10M_{\odot}$, $L=0.5L_{\rm Edd}$ and $a=0$. 
As expected, PD exhibits a maximum where the absorption opacity becomes comparable to the Thomson scattering one.

Exemplary results of the integrated disk spectra are shown in Figure~\ref{fig:disc_atm}, before and after accounting for relativistic effects.
The detailed modeling confirms the results from the simpler models: the polarization predicted by the standard thin disk model is lower than the observed one. 

\section{Polarization of the emission from slim and thick accretion discs}\label{section:slimdisc}
Here we investigate phenomenological thick disk models \citep{aWest} in order to check if the observed X-ray polarization properties can be explained in this scenario.
The models assume discs of various scale heights embedded in the Kerr space-time, with the disk matter orbiting the BH on GR Keplerian orbits. 
Following the classical treatment by \cite{Page_1974}, 
the disk material locally emits all the net energy that it gains by sinking towards the BH. 
Given the energy liberated in each radial bin per co-moving time, the temperature of the disk surface is calculated from the Stefan-Boltzmann law, accounting for the fact that a thicker disk has a larger surface area than a geometrically thin disc. It is furthermore assumed that the disk emits locally a modified BB energy spectrum adopting a constant spectral hardening factor of $1.8$.
The surface of the disk follows a simple phenomenological description inspired by the
Polish doughnut models by \citet{Abramowicz_1988}.
The discs are characterised by the scale height, $h/r$ with $h$ being the height of the disk where it is thickest, and $r$ being the corresponding disk radius.

\begin{figure}[h]
    \centering
    \includegraphics[height=8.5cm]{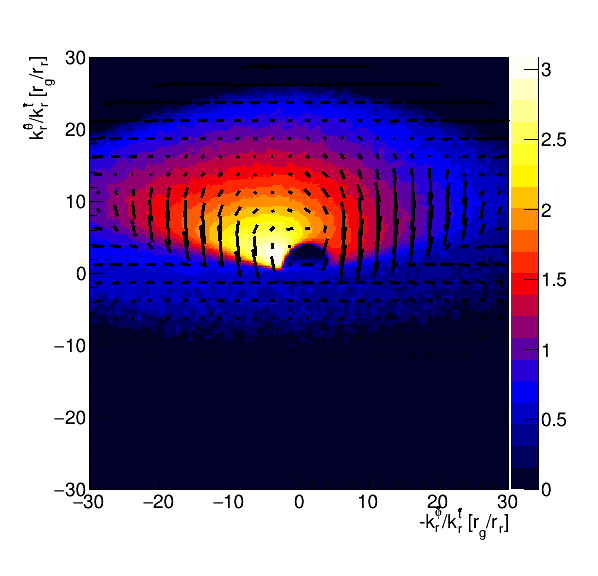}
    \includegraphics[height=8.5cm]{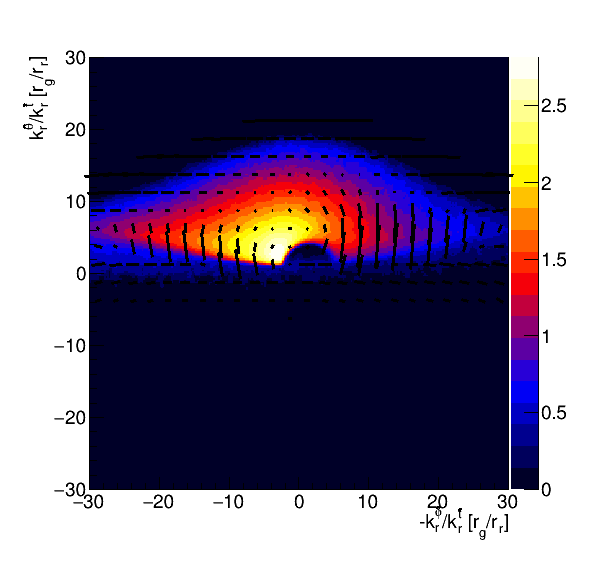}
    \caption{Total flux obtained from ray-traced simulations for a BH spin $a=0.98$, with $h/r = 0.5$ viewed at a $75\degr$ inclination (left) and $h/r = 0.3$ viewed at an $85\degr$ inclination (right).
    The colour scale shows the flux on a logarithmic scale in arbitrary units. Overlaid, the black vectors indicate the PA and have a length proportional to the PD, both as seen by a distant observer.}
    \label{fig:2Dthickdiscs}
\end{figure}

\begin{figure}[t]
    \centering
    \includegraphics[width=0.9\hsize]{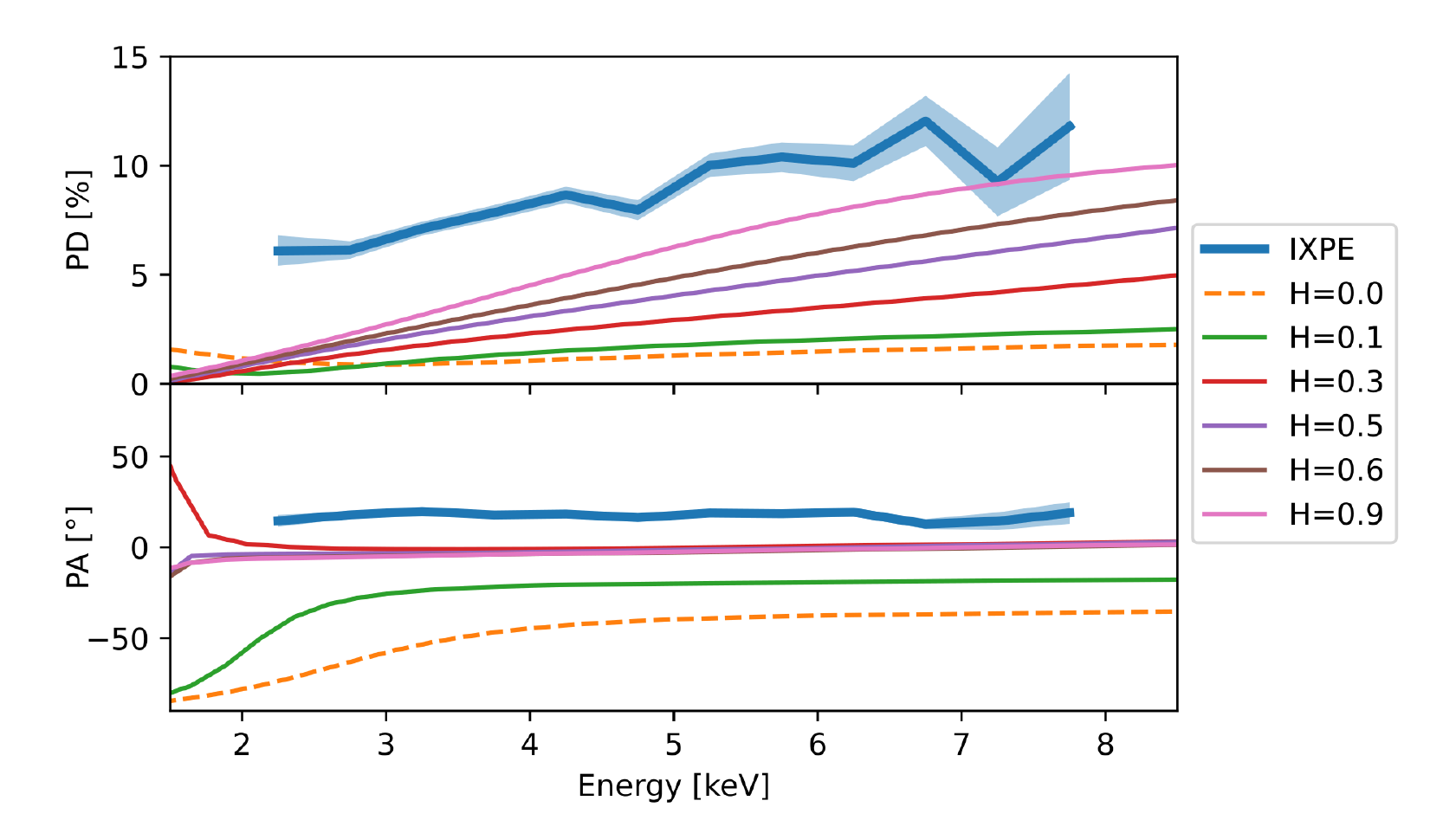}
    \caption{Polarization degree (top) and angle (bottom) for thin, slim, and thick discs for a BH spin of $0.98$ viewed at $i =  85\degr$, assuming a highly-ionized disk atmosphere. The model predictions are marked by solid lines of different colours (dashed lines correspond to the thin disk prediction), while {IXPE} data 
    are reported as a blue line in both panels. The shaded cyan region represents the errors at $68\%$ confidence level. 
    }
    \label{fig:thickdiscmodels}
\end{figure}

Figure~\ref{fig:2Dthickdiscs} shows the 2-D distribution of the surface brightness, PD and PA as seen by a distant observer. Both plots show the effect of the colder and X-ray dimmer outer portions of the slim discs shadowing the emission from the hot and bright inner regions.         
Figure~\ref{fig:thickdiscmodels} shows the PD and PA energy spectra for discs scale heights ranging between 0 and 0.9.
The PD exhibits an increasing behavior across the {IXPE} band, 
as the highest-energy emission from the innermost portion of the disk are most likely to scatter off the opposite side of the disc. On the other hand, the PA stays roughly constant with energy, except for very low scale heights. Note that since the orientation of the system is unknown, we can check only the trend in energy behavior of PA while its absolute value cannot be assessed. This is why we leave it unconstrained in the Figure~\ref{fig:thickdiscmodels}. The figure shows that increasing the scale height leads to overall higher values of the PD.
However, we could not find a parameter combination that achieves a PD as high as the observed ones. 

\begin{figure}[tbh]
    \centering
    \includegraphics[width=0.9\hsize]{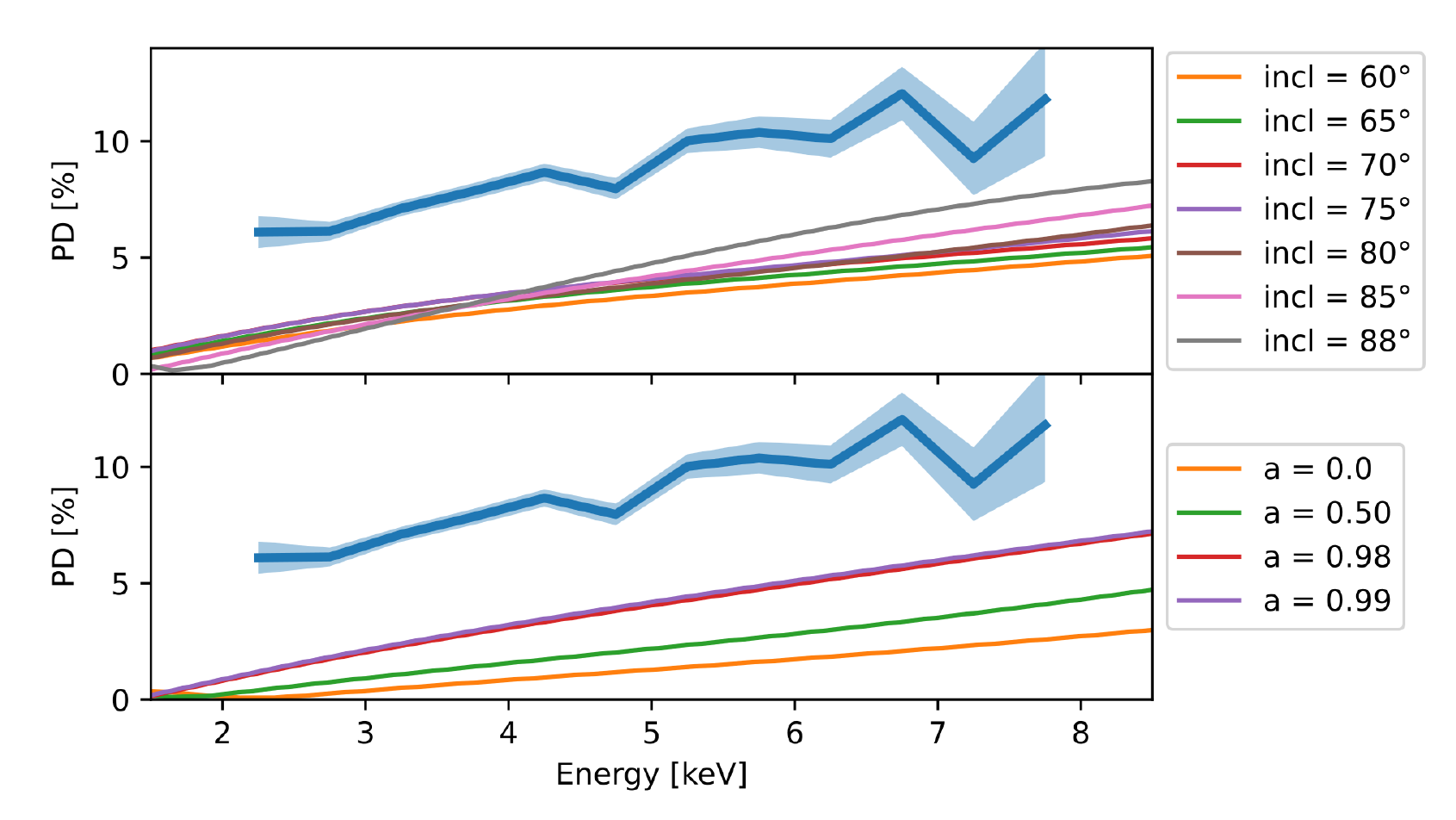}
    \caption{Polarization degree predictions for discs with scale height $0.5$. Top panel: Results for a BH with spin $a=0.99$, viewed at different inclination angles. Bottom panel: Results for a BH with different spins, viewed at an inclination of $85\degr$.}
    \label{fig:PDsthickdisc}
\end{figure}

Figure \ref{fig:PDsthickdisc} shows that the PDs increase with inclination and with the black hole spin. The dependence of the PD on inclination results from the combined effects of different emission angles of the photons reaching the observer, and different degrees of self-shadowing of the emission by thick discs.
Increasing the BH spin leads to higher PDs as higher spins correspond to the discs extending closer to the black holes, and leading to a higher fractions of photons reflecting off the discs.


\clearpage

\section{Additional figures}
\begin{figure}[thb]
    \hspace*{-1mm}\includegraphics[width=0.51\textwidth]{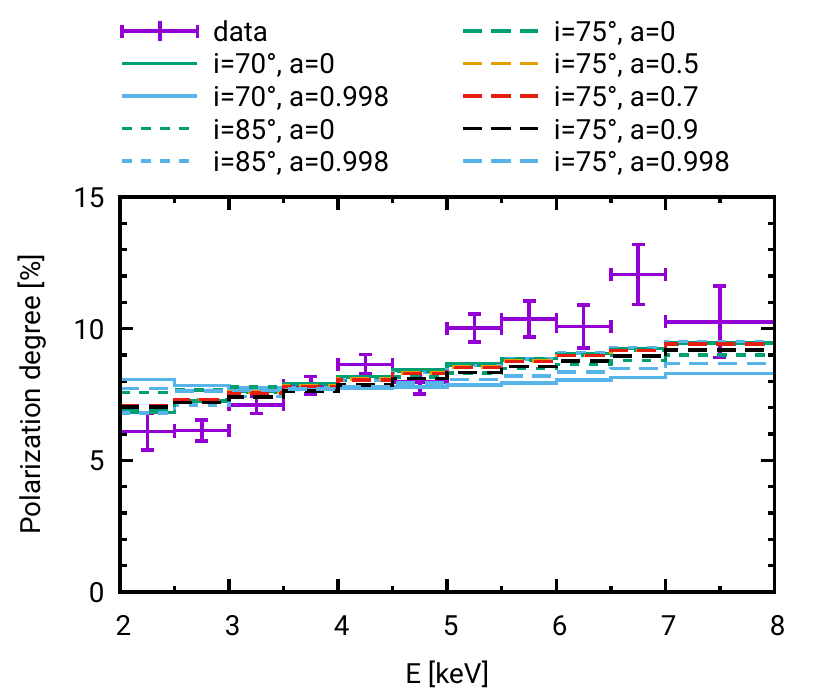}
    \hspace*{-1mm}\includegraphics[width=0.51\textwidth,trim={0 0 0 1.9cm},clip]{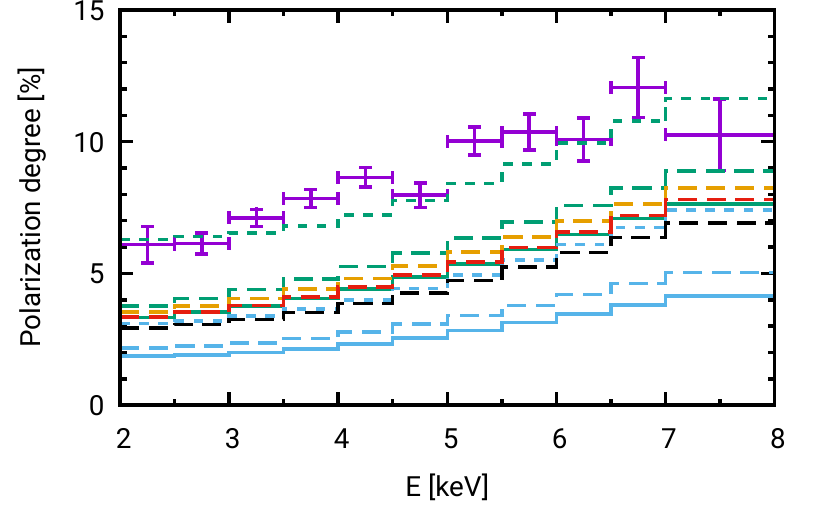}
    \caption{Attempted fits of the observed polarization degree with the models (B) (left) and (C) (right). 
    We did not account for self-irradiation in these computations.}
    \label{fig:model2}
\end{figure}

\begin{figure}[bht]
    \hspace*{-1mm}\includegraphics[width=0.51\textwidth]{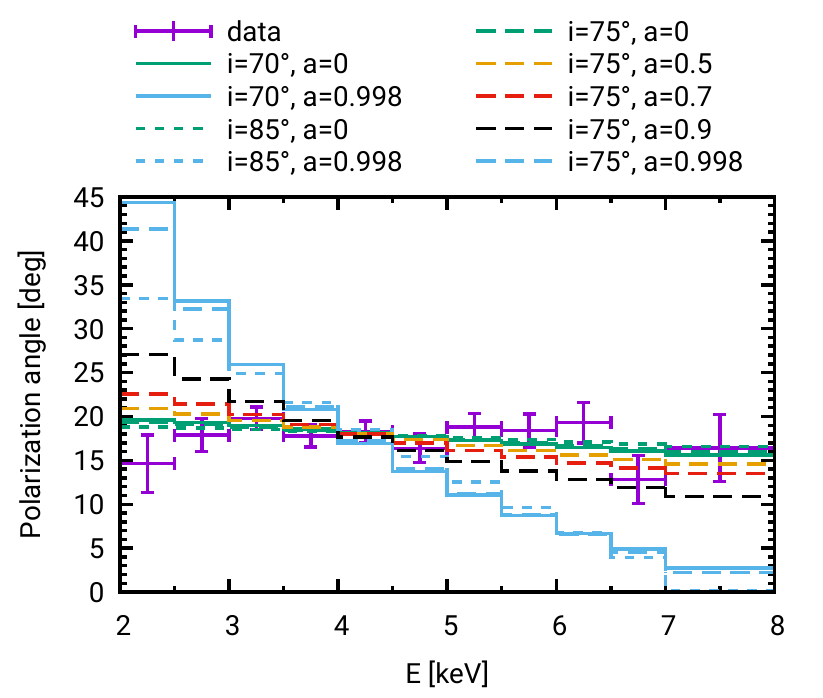}
    \hspace*{-1mm}\includegraphics[width=0.51\textwidth,trim={0 0 0 1.9cm},clip]{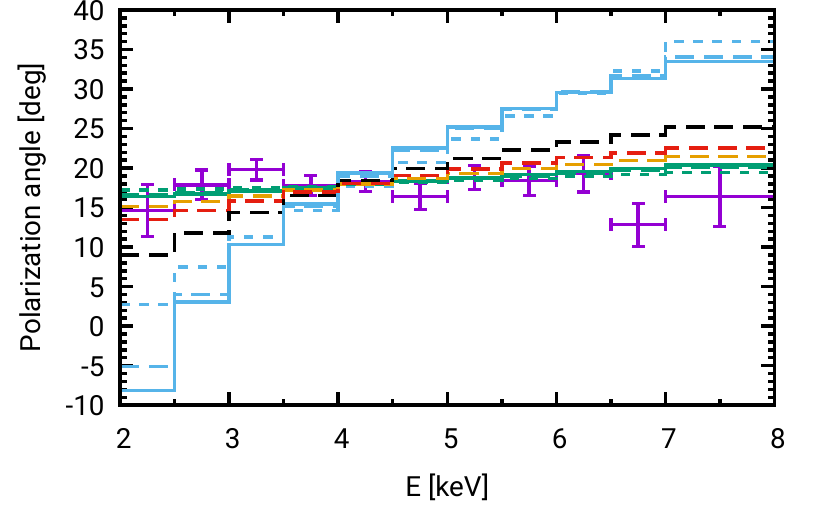}
    \caption{Polarization angle predicted by the models. 
The PA for both anticlock-wise (left) and clock-wise (right) directions of the system rotation are shown (disk and BH are co-rotating in both cases).
}
    \label{fig:pol_angle}
\end{figure}

\clearpage

\section{Additional tables}
\begin{table}[bht]
\centering
\caption{Comparison of the spectral fit results for different values of inclination and BH spin, with the {\sc kynbbrr} model representing the thermal component.} 
\begin{tabular}{cccccc}
\hline
\hline
\rule{0cm}{0.3cm}
Incl. & Spin & Mass & Accretion rate & Normalization & Fit goodness  \\  
\hline
\rule{0cm}{0.3cm}
 $i$ (deg) & $a$ & $M_{\textrm{bh}}$ ($M_\odot$) & $\dot{M}/\dot{M}_{\rm Edd}$ & norm & $\chi^2$ (792 dof) \\
\hline
$70$ & $0$     & $3$ (frozen)   & $0.451\pm0.003$   & $2.13\pm0.02\ ^{\rm(a)}$       & $730$ \\
     & $0.998$ & $23.59\pm0.10$ & $0.1221\pm0.0003$ & $0.75614$ (frozen) & $638$ \\
$75$ & $0$     & $3$ (frozen)   & $0.424\pm0.003$   & $3.02\pm0.03\ ^{\rm (b)}$       & $719$ \\
     & $0.5$   & $3.59\pm0.02$  & $0.1992\pm0.0005$ & $4.93827$ (frozen)  & $732$ \\
     & $0.7$   & $3.80\pm0.02$  & $0.1759\pm0.0005$ & $4.93827$ (frozen)  & $719$ \\
     & $0.9$   & $15.57\pm0.07$ & $0.2402\pm0.0006$ & $0.75614$ (frozen) & $670$ \\
     & $0.998$ & $29.5\pm0.1$   & $0.1005\pm0.0002$ & $0.75614$ (frozen) & $632$ \\
$85$ & $0$     & $3$ (frozen)   & $0.6336\pm0.0005$ & $4.93827$ (frozen)  & $728$ \\
     & $0.998$ & $48.7\pm0.2$   & $0.0724\pm0.0002$ & $0.75614$ (frozen) & $629$ \\
     \hline
\end{tabular}
\label{table:nicer_spinincl}
\begin{tablenotes}
\item Notes: If not noted otherwise,
a source distance of $11.5\,$kpc is assumed corresponding to a normalization of 0.76. 
In case the fitted BH mass hit the lower limit of $3\,M_\odot$, we assumed a distance of $4.5\,$kpc corresponding to a model normalization of 4.9. 
In case that the fitted BH mass hit the lower limit of $3\,M_\odot$ again, we assumed $3\,M_\odot$ as the BH mass and fitted the normalization which corresponds to ${\tt norm}=1/D^2_{\rm 10}$ with $D_{\rm 10}$ being the distance to the source in units of $10\,$kpc. The fitted values of {\sc tbabs} and {\sc cloudy} parameters were close to the values shown in Table~\ref{table:nicernustar}. \\ 
\item {$^{\rm (a)}$} Corresponding to the distance of $6.85$\,kpc.
\item {$^{\rm (b)}$} Corresponding to the distance of $5.75$\,kpc.
\end{tablenotes}
\end{table}

\begin{table}[tbh]
\centering
\caption{Comparison of 
PD and PA fits using different flavors of the {\sc kynbbrr} model. Note that both PD and PA were binned in 11 energy bins. The fits with $\chi^2_{\rm tot}<22$ are denoted in bold and the best fit parameter values of these cases are shown in Table \ref{table:ixpe_params}.} 
\begin{tabular}{cccccc}
\hline
\hline
\rule{0cm}{0.3cm}
Incl. & Spin & model (A) & model (B) & model (C) & model (D) \\  
\hline
\rule{0cm}{0.3cm}
 $i$ (deg) & $a$ & $\chi^2({\rm PD/PA/tot})$ & $\chi^2({\rm PD/PA/tot})$ & $\chi^2({\rm PD/PA/tot})$ & $\chi^2({\rm PD/PA/tot})$ \\
\hline
$70$ & $0$     & 1573/8.9/1581 & 33/8.0/40  & 626/7.7/634  & {\bf 11.5/7.9/19.4} \\
     & $0.998$ & 2236/280/2517 & 84/80/164  & 1551/66/1617 & 12.2/57/69 \\
$75$ & $0$     & 1132/7.8/1140 & 35/7.8/43  & 387/7.7/395  &  {\bf 11.8/7.8/19.6} \\
     & $0.5$   & 1529/13/1542  & 37/8.8/45  & 506/8.3/514  &  {\bf 12.2/8.3/20.5}\\
     & $0.7$   & 1722/21/1743  & 37/10/47   & 606/9.8/616  &  {\bf 12.5/9.3/21.8} \\
     & $0.9$   & 2077/53/2130  & 41/15/56   & 828/15/843   & 12.7/12.9/25.7 \\
     & $0.998$ & 2162/258/2420 & 65/56/121  & 1309/48/1357 & 12.1/41/53 \\
$85$ & $0$     & 658/7.8/666   & 55/7.7/63  & 42/7.9/50    &  {\bf 13.3/7.9/21.2} \\
     & $0.998$ & 2098/187/2285 & 30/21/51   & 765/21/786   & 12.9/18.6/31.54 \\
\hline
\end{tabular}
\label{table:ixpe_chi2}
\end{table}

\begin{table}[tbh]
\centering
\caption{Best fit parameters of the {\sc kynbbrr} model (D) with outflowing ionized layer. Since the optical thickness of the layer and its outflow speed were degenerate and are anti-correlated, we eventually froze the optical thickness at $\tau=7$. The outflow speed (last column) is characterized by its value at the radius where the disk temperature peaks, $\beta(T_{\rm max})$.}
\begin{tabular}{cccccc}
\hline
\hline
Incl. & Spin & orientation & speed norm & speed index & speed \\  
\hline
\rule{0cm}{0.3cm}
 $i$ (deg) & $a$ & $\chi_{\rm o}$ & $\beta_0$ & $q$ & $\beta(T_{\rm max}$)\\
\hline
\hline
$70$ & $0$     & $-70.7\pm0.5$ & $0.65\pm0.14$ & $0.54\pm0.19$ & 0.50\\
$75$ & $0$     & $-71.2\pm0.5$ & $0.56\pm0.18$ & $0.72\pm0.29$ & 0.40\\
     & $0.5$   & $-70.1\pm0.5$ & $0.65\pm0.18$ & $0.73\pm0.26$ & 0.47\\
     & $0.7$   & $-68.9\pm0.5$ & $0.71\pm0.17$ & $0.70\pm0.23$ & 0.53\\
$85$ & $0$     & $-72.5\pm0.5$ & $0.6$ (frozen) & $2.1\pm0.2$ & 0.22\\
\hline
\end{tabular}
\label{table:ixpe_params}
\end{table}

\clearpage

\bibliography{4U1630}{}
\bibliographystyle{aasjournal}

\end{document}